\documentclass[12pt]{article}
\usepackage{amsmath}
\usepackage{amssymb}
\usepackage{bbm}
\usepackage{color}
\usepackage{graphicx,psfrag,epsf,epsfig,epstopdf}
\usepackage{enumerate}
\usepackage{threeparttablex}
\usepackage{adjustbox}
\usepackage{natbib}
\usepackage{multirow}
\usepackage{colortbl}
\usepackage{xcolor}
\usepackage{url} 
\usepackage{ulem,xpatch}
\usepackage{hyperref} 

\newtheorem{assumption}{Assumption}

\graphicspath{{figures/}}
\newtheorem{example}{Example}[section]

\newcommand{\blind}{0}

\begin{document}

\def\spacingset#1{\renewcommand{\baselinestretch}%
{#1}\small\normalsize} \spacingset{1}


\if0\blind
{
  \title{\bf Group Sequential Design for Non-Proportional Hazards: Logrank, Weighted Logrank, and MaxCombo Methods}
  \author{Yujie Zhao\textsuperscript{1}, 
          Yilong Zhang\textsuperscript{2},  
          Larry Leon\textsuperscript{1}, 
          Keaven Anderson\textsuperscript{1}\\
    \textsuperscript{1} Merck \& Co., Inc., Rahway, NJ, USA\\
    \textsuperscript{2} Meta Platforms Inc., Menlo Park, CA, USA }
  \maketitle
} \fi

\if1\blind
{
  \bigskip
  \bigskip
  \bigskip
  \begin{center}
    {\LARGE\bf Group Sequential Design for Non-Proportional Hazards: Logrank, Weighted Logrank, and MaxCombo Methods}
\end{center}
  \medskip
} \fi

\bigskip
\begin{abstract}
Non-proportional hazards (NPH) are often observed in clinical trials with time-to-event endpoints. A common example is a long-term clinical trial with a delayed treatment effect studying immunotherapy for cancer. When designing clinical trials with time-to-event endpoints, it is crucial to consider NPH scenarios to gain a complete understanding of design operating characteristics. In this paper, we focus on group sequential design for three NPH methods: the logrank test, the weighted logrank test, and the MaxCombo combination test. For each of these approaches, we provide analytic forms of design characteristics that facilitate sample size calculation and bound derivation for group sequential designs. Examples are provided to illustrate the proposed methods. To facilitate statisticians in designing and comparing group sequential designs under NPH, we have implemented the group sequential design methodology in the \textit{gsDesign2} R package at \url{https://cran.r-project.org/web/packages/gsDesign2/}. 
\end{abstract}

\noindent%
\vfill

\newpage
\spacingset{1.45} 


\section{Introduction}
\label{sec:introduction}

In clinical trials with time-to-event endpoints, non-proportional hazards (NPH) are frequently observed. A notable example in oncology is the immune-directed anticancer therapies \citep{reck2016pembrolizumab}, which activate the immune system to induce an anti-tumor response, potentially leading to delayed treatment effects \citep{mick2015statistical}. Other examples of NPH include the crossing survival curve pattern and the strong null scenario, where the control therapy shows better outcomes early on and converges afterward, while the experimental therapy is never superior to the control \citep{wassie2023delayed}.

When designing a clinical trial under NPH, two major challenges arise. First, it is important to explore alternative approaches to quantify treatment differences, beyond the commonly used logrank test. This is primarily because the logrank test may demonstrate reduced power under NPH compared to proportional hazards (PH) \citep{leon2020weighted, mukhopadhyay2020statistical}. Second, a fixed design with a single analysis may not allow for early termination if the study has sufficient evidence of treatment effect. As a result, the utilization of group sequential designs has become more prevalent. It incorporates interim analyses (IAs), which permit multiple assessments of the data before the study is concluded, thereby facilitating early termination when adequate evidence is obtained (see a brief introduction to group sequential designs in Appendix \ref{appendix: into of gsd}).

In this paper, we investigate three NPH hypothesis testing procedures and their corresponding design derivations in group sequential designs: (M.1) the logrank test utilizing the AHR method, (M.2) the weighted logrank test, and (M.3) the MaxCombo test.

\textbf{(M.1) Logrank test (LR) using the AHR method (AHR).} The hazard ratio (HR) is a widely used metric for assessing treatment effects in survival analysis. 
Under NPH, the natural extension to average hazard ratio (AHR) has been recommended \citep{AHRSchemper2009, kalbfleisch1981estimation}.
Although various AHR definitions have been suggested by \cite{AHRSchemper2009}, 
our focus is on the approach proposed by \cite{mukhopadhyay2020statistical} as it aligns with the widely used logrank test and Cox model estimation.
We expand upon this approach by using a piecewise enrollment and piecewise proportional hazards model, and we also refine the asymptotic theory for group sequential design.

\textbf{(M.2) Weighted logrank (WLR) test.} In the context of NPH, researchers have investigated both the analysis and study design problems associated with the WLR test, which has the potential to improve power or reduce sample size \citep{luo2019design}. In the framework of WLR test, one key issue is the selection of time-dependent weights. \cite{harrington1982class} used weight functions based on survival functions or at-risk proportions \citep{ tarone1977distribution}. In recent years, \cite{Magirr2019, magirr2021non} proposed a modestly weighted logrank test that avoids the issue of near-zero weighting for early observations, which can inflate Type I error. 
Another key issue is the asymptotical theory. \cite{tsiatis1982repeated} proved that weighted logrank group sequential tests asymptotically follow the multivariate normal distribution based on independent increments. While this is normally applied to proportional hazard scenarios, the theory applied to non-proportional hazard scenarios as well. Thus, we focus here on computing statistical information and the expected value of Z-tests under non-proportional hazard models.

\textbf{(M.3) MaxCombo Test.} Recent work in reviewing the WLR test has revealed that weight selection can be sensitive in different scenarios \citep{roychoudhury2021robust}. To address this challenge, researchers have proposed a versatile MaxCombo test originally proposed by \cite{lee1996some}. The MaxCombo test mathematically selects the maximum value of a set of different WLR tests, each of which is designed to be powerful in detecting a specific NPH or PH pattern. As a result, the MaxCombo test can yield competitive and robust power, which is quite close to optimal across many scenarios, irrespective of whether it is PH or NPH.  
For a fixed time point, \cite{Karrison2016} found the joint distribution of multiple WLR tests with Fleming-Harrington weights. In this paper, we extend distributional calculations to group sequential testing, noting that methods other than the canonical group sequential asymptotic model are required.

While reviewing the existing NPH literature, we acknowledge that work has been done for the aforementioned three tests in trial designs, based on either simulation-based or analytical methodologies. However, there are certain limitations to existing work. For example, \cite{yung2020sample} investigate the WLR test in fixed design. Their methodologies have not extended to group sequential designs. \cite{wang2021simulation} provides an analytical form for group sequential design for the MaxCombo test, but does not cover the LR test.  \cite{luo2019design} discuss the logrank statistics and its variance-covariance structure under NPH, yet, many design operating characteristics (such as the boundaries and crossing probability) are not discussed. \cite{roychoudhury2021robust} discuss the MaxCombo test to provide robust power based on simulations without an analytic form of design characteristic. \cite{bautista2021sample} analyzes the sample size and power under multiple case studies, but the analytic forms are not presented. Our previous work \cite{zhao2024group} offers a high-level overview of three testing methods with examples. While it presents the asymptotic theory of the test statistics, it does not include the derivation of design characteristics using this asymptotic theory.  Additionally, the validation of this asymptotic theory was not covered, which we aim to address in this paper. Furthermore, we establish the connection between the AHR method and the WLR test, which was not addressed in our previous work \cite{zhao2024group}.

The objective of this paper is to build on the existing literature to further enhance the aforementioned three tests.
First, we thoroughly examine and present an analytical formulation for group sequential design utilizing the LR test. 
Second, we derive analytical forms of group sequential design with a flexible choice of spending functions. Additionally, we establish both the efficacy and futility bounds in a canonical form for LR and WLR tests, as well as the noncanonical form for the MaxCombo test. To facilitate statisticians in designing and comparing group sequential designs under NPH, we implement the group sequential design methodology with multiple examples in the \cite{gsDesign2}. 
Finally, we establish the asymptotic equivalence between the AHR derived from the LR test and the one derived from the WLR test under a piecewise model.

The remainder of this paper is organized as follows. In Section \ref{sec: method}, we introduce the test statistics of the three NPH methods. In Section \ref{sec: design char}, we present related design characteristics, including boundaries, crossing probabilities, sample size, and number of events. In Section \ref{sec: simulation}, we present a simulation to verify the asymptotic theory proposed in Section \ref{sec: method}. In Section \ref{sec: application}, we apply the three approaches in a case study. A brief discussion is provided in Section \ref{sec: discussion}.

\section{Test statistics}
\label{sec: method}


We focus on study designs involving two treatment groups
with $n$ planned subjects and $K$ analyses at calendar time of $\tau_1, \ldots, \tau_K$ with information fraction of $t_1, \ldots, t_K$. At the $k$-th analysis, there are $n_k$ subjects included. 
Subjects are enrolled for an accrual duration of $\tau_a$ and followed for an additional period of $\tau_f$,
resulting in a total study duration of $\tau = \tau_a + \tau_f$.
For the $i$-th subject, we denote $X_i, R_i, T_i, L_i$ as the treatment assignment ($X_i = 0$ for control arm and $X_i = 1$ for treatment arm), time of study entry, time from study entry until an event occurs, and time from study entry until lost follow-up, respectively. Additionally, we denote $U_{i,k} = \min(T_i, L_i, \tau_k - R_i), C_{i,k} = \min(L_i, \tau_k - R_i)$ and $\delta_{i,k} = \mathbbm 1(T_i=U_{i,k})$ as the observed survival time, observed censoring time, and indicator of events of subject $i$ at analysis $k$. We further assume:

\begin{assumption}
\label{assump: common across methods}
    Suppose within each treatment group, the following two conditions hold.
    \begin{enumerate}   
      \item For treatment group $j (j=0,1)$, survival times are independent and identically distributed with cumulative distribution function (cdf) 
      $P(T_i\le t) = F_j(t)$,  
      probability density function (pdf)  
      $f_j(t)=dF_j(t)/dt$, 
      and hazard rate $\lambda_j(t) = f_j(t)/(1-F_j(t)).$   
      \item Loss-to-follow-up times $L_i$ are independent and identically distributed within each treatment group and are independent of $T_i$. 
   \end{enumerate}
\end{assumption}


\subsection{LR test using the AHR method}
\label{sec: test -- ahr}

When PH are assumed, the exponential survival distribution is a parametric model often used for sample size computation. 
To accommodate NPH, the AHR method uses a piecewise model with changing hazard rates over time:

\begin{assumption}
\label{assump: ahr piecewise}
   The AHR method for the logrank test assumes a piecewise model:
   \begin{enumerate}
       \item The enrollment rate is piecewise constant. 
       That is, we assume subjects enroll according to a Poisson process with an entry rate $g(u) \geq 0$ for $u \geq 0$.
       Thus, the expected number of subjects enrolled by study time $t$ is simply $G(t) = \int_0^t g(u)du$.
       We note that \cite{luo2019design} used a piecewise uniform distribution with a fixed maximum duration, slightly different than here where there is only an expected trial duration, but the actual duration may be shorter or longer.
       
       \item The dropout rate is piecewise constant, i.e.,
       it equals to $\eta_i \geq 0$ in the $i$-th interval. 
       This may vary by treatment group, but for the implementation of the AHR method, the rates are the same for each treatment group.
       
      \item The time-to-event rates are piecewise constant, i.e., 
      it equals $\lambda_j \ge 0$ in the $j$-th interval.
      We constrain that at in least one interval, we have $\lambda_m > 0$.
   \end{enumerate}
\end{assumption}

The three piecewise assumptions divides the entire timeline into $M$ intervals, where within each interval, the enrollment, dropout rates, and time-to-event rates remain constant. An example can be found in \cite{zhao2024group}. These assumptions offer a flexible approximation method suitable for a wide range of design scenarios. These three piecewise assumptions provide a flexible approximation method for a broad set of design scenarios. 
Furthermore, these assumptions are easy to explain to collaborators.
The first assumption tailors the case where the enrollment rate is anticipated to change over time. 
The last two assumptions offer flexibility when dropout or failure rates change over time; allowing a failure rate of 0 enables a fixed follow-up duration for each study participant. 

Note that these piecewise assumptions are an extension of exponential failure and dropout rates along with proportional hazards of \cite{LachinFoulkes}. This offers flexibility to fit cases with NPH such as a delay in treatment effect in Section \ref{sec:introduction}. Our approach to sample size approximation is inspired by \cite{LachinFoulkes}.
In our case, rather than a single HR, we have a finite set of HRs in the piecewise model. As in \cite{LachinFoulkes}, we look at the parametric estimation initially and then extend this from a single HR across all piecewise intervals to a weighted average of the log-hazard ratio across intervals with weighting by the expected Fisher information under the null hypothesis under the piecewise parametric model. This is done with the following steps:

\textbf{\underline{Step 1:}} Write a likelihood for each piecewise interval and  compute the statistical information for the logarithm of the HR in that interval. This is simply based on the expected number of events in each arm; see Appendix \ref{appendix: proof -- ahr -- ahr asy dist}. 

\textbf{\underline{Step 2:}} Compute the average HR as the exponent of the weighted average of the piecewise assumed logarithms of the HRs for each piecewise interval. Weight according to the statistical information (expected events) under the null hypothesis in each piecewise interval. This is to minimize the variance among weighted averages of the log HRs from Step 1. The resulting statistical information (inverse variance) under a local alternative approach \citep{schoenfeld1981asymptotic, luo2019design} is proportional to the total expected number of events. 

\textbf{\underline{Step 3:}} Assume a Wald-like Z-test based on the AHR model: estimated treatment effect based on the piecewise exponential model. This involves the expected value of log(average HR) and its variance approximation from Step 2.

\textbf{\underline{Step 4:}} As part of the \cite{LachinFoulkes} method, statistical information is computed under both the null and alternative hypotheses. Any efficacy bound is computed under the null hypothesis to control Type I error. Futility bounds and sample size are generally computed under the alternative hypothesis.

While we could take a full likelihood approach, the above approach avoids the computation of second partials of the log-likelihood of all piecewise HRs and hazard rates.
Note that all of this has been extended to stratified populations in both the \textit{gsDesign2} and \textit{simtrial} R packages.
The test statistic used is the Z-value version of the logrank test; this is the unweighted version of the WLR test in Section \ref{sec: test -- wlr}, whose 
canonical joint distribution is:
\begin{itemize}
    \item $Z_1^{\text{(lr)}}, Z_2^{\text{(lr)}}, \ldots, Z_K^{\text{(lr)}}$ have a multivariate normal distribution.
    \item $E(Z_k^{\text{(lr)}}) = 0$ under the null hypothesis.
    \item $\text{Cov}(Z_i^{\text{(lr)}}, Z_j^{\text{(lr)}}) = \sqrt{t_i/t_j}$ for any $1 \leq i \leq j \leq K$ under the null hypothesis.
\end{itemize}

Under the alternate hypothesis, if we denote the treatment effect as $\theta_k$ and statistical information under null hypothesis as $\mathcal I_{k, H_0}$, then we have the asymptotic mean and variance of $Z_k^{\text{(lr)}}$ as $\theta_k \sqrt{\mathcal I_{k, H_0}}$ and $\mathcal I_{k, H_1} / \mathcal I_{k, H_0}$, respectively. And the covariance is $\text{Cov}(Z_i^{\text{(lr)}}, Z_j^{\text{(lr)}}) = \frac{1}{\sqrt{t_i t_j}}\text{Cov}(B_i, B_j) = \frac{1}{\sqrt{t_i t_j}} \text{Var}(B_i) = \sqrt{\frac{t_i}{t_j}} \frac{\mathcal I_{i, H_0}}{\mathcal I_{i, H_1}}$ under the alternate hypotheses. When the local alternative assumption is satisfied (see Appendix \ref{appendix: local alternative}), we have $\text{Cov}(Z_i^{\text{(lr)}}, Z_j^{\text{(lr)}}) \approx \sqrt{t_i/t_j}$, which is in the format of the canonical joint distribution introduced in Section \ref{sec:introduction}.

\subsection{WLR test}
\label{sec: test -- wlr}

The purpose of the WLR test is to compare the survival curves of two groups. In this scenario, the null hypothesis is stated as $H_0: \overline F_0(\cdot) = \overline F_1(\cdot)$, where $\overline F_j(\cdot)$ represents the complement of the cumulative distribution function (cdf) of the survival distribution for group $j \in \{0, 1\}$.

\begin{assumption}
\label{assump: wlr}
   The WLR method is based on certain assumptions:
   \begin{enumerate}
       \item the time of study entry $\{R_i\}_{i = 1, \ldots, n}$ has continuous cdf denoted as $H(\cdot)$;
       \item the time to loss follow-up in group $j$, i.e., $\{L_i | X_i = j\}_{i \in \{X_i = j\}}$ has continuous cdf $G_j(\cdot)$.
   \end{enumerate}
\end{assumption}

Under the aforementioned assumptions, the WLR method employs the weighted logrank test, which is an extension of the logrank test that incorporates weights to examine the null hypothesis $H_0$.
The test statistics for the $k$-th analysis are as follows:
\begin{equation}
  Z_k^{(\text{wlr})}
  =
  \frac{U_k}{\sqrt{V_k}}
  =
  \frac{
  \sum\limits_{\{s:\; s < \tau_k\}}
  a(s) 
  \left(
    X_{(s)} - \frac{\overline Y_1(s)}{\overline Y_0(s) + \overline Y_1(s)}
  \right)
  }
  {
  \sqrt{
  \sum\limits_{\{s:\; s < \tau_k\}}
  a(s)^2
  \frac{
  \overline Y_0(s)
  \overline Y_1(s)}{\left[\overline Y_0(s) + \overline Y_1(s)\right]^2}
  }
  }.
  \label{eq:WlrZDef}
\end{equation}
Here ${\{s:\; s < \tau_k\}}$ is the complete set of event times before the $k$-th analysis. The $X_{(s)}$ is the assigned treatment for the subject failing at time $s$, and the $\overline{Y}_j(s)$ is the number of at-risk subjects in group $j$ at time $s$.




The numerator $U_k$ in \eqref{eq:WlrZDef} might not correspond to a measure of treatment efficacy. However, with some linear transformation, we can show $U_k$ as a weighted summation of the difference in estimated hazards in discrete time:
$
  U_k
  =
  \sum\limits_{\{s:\; s < \tau_k\}}
  a(s) 
  \frac{
    \overline{Y}_0(s)
    \overline{Y}_1(s)
  }{
    \overline{Y}_0(s)
    + 
    \overline{Y}_1(s)
  }
  \left(
    \frac{X_{(s)}}{\overline{Y}_1(s)} 
    -
    \frac{1 - X_{(s)}}{\overline{Y}_0(s)} 
  \right).
$
Following Appendix A.3 in \cite{yung2020sample}, one has
\begin{equation}
  \sqrt{n_k}
  \left( U_k/n_k - \Delta_k \right)
  \overset{d}{\to}
  N(0, \widetilde\sigma^2_{b,k}),
  \label{eq:WlrUDist}
\end{equation}
where $\overset{d}{\to}$ represents convergence in distribution,  $\widetilde\sigma^2_{b,k}$ is a constant, and
\begin{equation}
  \Delta_k
  =
  \int_{0}^{\tau_k}
  w(s)
  \frac{
    p_{0,k} \pi_{0,k}(s) \;
    p_{1,k} \pi_{1,k}(s)
  }{
    \pi_k(s)
  }
  \left[\lambda_{1}(s)-\lambda_{0}(s)\right]
  ds.
  \label{eq:WlrDletaK}
\end{equation}
Here the term $p_{j,k}$ is the randomization probability of group $j$ at $k$-th analysis, i.e., $p_{j,k} = n_{j,k}/(n_{0,k} + n_{1,k})$. In addition, $\pi_{j,k}(t)$ is the expected at-risk probability of group $j$, i.e., 
$
  \pi_j(t) 
  \triangleq
  E\left[\mathbbm 1\{U_i \geq t\} | X_i = j\right]
  =
  \overline{F}_j(t) \overline{G}_j(t) H(\min\{\tau_a, \tau - t\}).
$
And $\pi_k(t) = p_{0,k} \pi_{0,k}(t) + p_{1,k} \pi_{1,k}(t)$ is the overall at-risk probability. 
The function $w(t) = \lim_{n \to \infty} a(t)$ denotes the limit of $a(t)$, which weights the different hazard ratios over time.  In the literature, one of the popular weight functions is the Fleming-Harrington (FH) test 
$
  w(t) = 
  \left[\bar F(t-) \right]^p \left[ 1 - \bar F(t-) \right]^q
$
where $p \geq 0, q \geq 0$ and $\bar F(t-)$ is the left-continuous version of the Kaplan-Meier estimator for the pooled sample (see Section 2 in \cite{leon2020weighted}). 
Another commonly used weight function is the Magirr-Burman weight \cite{Magirr2019}: $w(t) = 1/\bar F(\min\{t, t^*\})$. Initially, the weight begins around 1 for the first event and gradually increases until time $t^*$. However, beyond time $t^*$, it opts to maintain the weights at the largest value achieved before $t^*$. In their investigation, \cite{Magirr2019} also explore the issue of type I error inflation. They find that if the scores are not non-increasing, the type I error can be inflated under the strong null scenario. Therefore, when the strong null scenario is likely to occur, users are advised to thoroughly examine the scores during the weight selection process. In a subsequent publication by \citep{magirr2021non}, this weight function has been generalized as $w(t) = \min\{w_{\max}, 1/\bar F(\min(t, t^*)\}$, where the maximal weight is capped at $w_{\max}$.  \cite{xu2017designing} proposes to have $w(t) = 0$ when there is a delayed treatment effect and $w(t) = 1$ afterward, which takes into account only the events accumulated after the delayed effect.

From \cite{yung2020sample} the denominator in \eqref{eq:WlrZDef}, $V_k$ has
\begin{equation}
  V_k/n_k \overset{p}{\to} \sigma_k^2.
  \label{eq:WlrVDist}
\end{equation}
Here, $\overset{p}{\to}$ represents convergence in probability. The values of $\sigma_k$ vary depending on whether it is under the null hypothesis or the alternate hypotheses:
\begin{equation}
\label{equ: sigma of wlr}
  \left\{
  \begin{array}{ccl}
     \sigma_k^{2} | H_1
  & = &
  \int_{0}^{t_k}
  w(s)^{2}
  \frac{
    p_{0,k}\pi_{0,k}(s) \;p_{1,k}\pi_{1,k}(s)
  }{
    \left[p_{0,k} \pi_{0,k}(s) + p_{1,k} \pi_{1,k}(s)\right]^{2}
  }dv(s) \\
  \sigma_k^{2} | H_0
  & = &
  \int_{0}^{t_k}
  w(s)^{2}
  \frac{
    p_{0,k} \pi_{0,k}(s) \; p_{0,k} \pi_{0,k}(s)
  }{
    \left[p_{0,k} \pi_{0,k}(s) + p_{0,k} \pi_{0,k}(s)\right]^{2}
  }dv(s)
  =
  \int_{0}^{t_k}
  w(s)^{2}\;
  \frac{
    p_{0,k} \pi_{0,k}(s)
  }{
    2
  }dv(s)
  \end{array}
  \right.,
\end{equation}
where  $v(t) = p_0 v_0(t) + p_1 v_1(t)$ is the failure probability with $v_j(t)$ representing the probability that a subject in group $j$ will experience an event within time $t$, i.e.,
$
  v_j(t) 
  \triangleq 
  E\left[\mathbbm 1\{U_i \leq t, \delta_i = 1\} | X_i = j\right]
  =       
  \int_0^t f_j(s) \overline{G}_j(s) H(\min\{\tau_a, \tau - s\}) ds.
$

Combining the results from \eqref{eq:WlrUDist} and \eqref{eq:WlrVDist}, we show $Z_k^{\text{(wlr)}}$ 
has the canonical joint distribution \citep{wang2021simulation}: 
\begin{itemize}
    \item $Z_1^{\text{(wlr)}}, Z_2^{\text{(wlr)}}, \ldots, Z_K^{\text{(wlr)}}$ have a multivariate normal distribution.
    \item $E(Z_k^{\text{(wlr)}}) = 0$ under the null hypothesis.
    \item $\text{Cov}(Z_i^{\text{(wlr)}}, Z_j^{\text{(wlr)}}) = \sqrt{\sigma_i/\sigma_j}$ for any $1 \leq i \leq j \leq K$ under the null hypothesis.
\end{itemize}
Under the alternate hypotheses, we have $E(Z_k^{\text{(wlr)}}) = \sqrt{n_k} \Delta_k / \sigma_k$. 
When the local alternative assumption holds (see Appendix \ref{appendix: local alternative}), one can get the asymptotic variance of $Z_k^{(\text{wlr})}$ as 1 under both the null and alternative hypotheses.


\subsection{MaxCombo test}
\label{sec: test -- mc}
The MaxCombo test \citep{lee1996some} considers the maximum value obtained from a combination of $L$ WLR tests, enabling robust test sensitivity under a variety of scenarios:
\begin{equation}
  Z_k^{(\text{mc})}
  =
  \max
  \left\{
    Z_k^{(\text{wlr}_1)},
    Z_k^{(\text{wlr}_2)}, 
    \ldots,
    Z_k^{(\text{wlr}_L)}
  \right\},
  \label{eq:ComboGDef}
\end{equation}
where $Z_k^{(\text{wlr}_i)}$ is a test statistic from a weighted logrank test.
Considering $Z_k^{(\text{mc})}$ involves a maximum operator, there is no canonical joint multivariate normal
distribution. However, the asymptotic normal distribution of $Z_k^{(\text{wlr}_i)}$ can be used to derive the type I error, power, and other design characteristics. 
In Appendix \ref{appendix: examples to solve upper and lower bounds}, we provide an example showing the calculation of the crossing probabilities in the MaxCombo test using the WLR test, where the crucial aspect lies in determining the asymptotic distribution of the set $\{ Z_k^{(\text{wlr}_i)}\}_{k=1, \ldots, K;\; i = 1, \ldots, L}$. 
This task requires the covariance matrix of the MaxCombo test. As shown in \cite{kundu2023closed}, there is an analytical formula for the covariance matrix under no-censoring assumption. In this paper, we derive the covariance matrix by removing this assumption. Without loss of generality, we take the WLR test with FH weighting as an illustrative example to show the correlation structure.

The first type of correlation is the correlation within the analysis between different tests. In the context of a fixed analysis $k$, the correlation between two WLR tests with FH weights of FH$(p_i, q_i)$ and FH$(p_j, q_j)$ is represented as
\begin{eqnarray*}
  \text{Corr}
  \left( 
    Z_k^{(\text{wlr}_i)}, 
    Z_k^{(\text{wlr}_j)}
  \right) 
  =
    \text{Var}
    \left(
      U_k^{(\text{wlr}_{ij})}
    \right)
    \bigg /
    \sqrt{
    \text{Var}
    \left( U_k^{(\text{wlr}_i)} \right)
    \text{Var}
    \left( U_k^{(\text{wlr}_j)} \right)
    },
\end{eqnarray*}
where $U_k^{(\text{wlr}_i)}, U_k^{(\text{wlr}_i)}, U_k^{(\text{wlr}_{ij})}$ are the numerator of the WLR test statistics with the weights of FH$(p_i, q_j)$, FH$(p_j, q_j)$ and FH$((p_i + p_j)/2, \; (q_i + q_j)/2)$. 

The second type of correlation is the within-test correlation between different analyses. Under the fixed WLR test with the weight of FH$(p_i, q_i)$, the correlation between the $k_1, k_2$-th analysis ($1 \leq k_1 \leq k_2 \leq K$) is 
\begin{equation*}
  \text{Corr}
  \left(
    Z_{k_1}^{(\text{wlr}_i)}, 
    Z_{k_2}^{(\text{wlr}_i)}
  \right)
  =
  \sqrt{
    \text{Var}\left(U_{k_1}^{(\text{wlr}_i)} \right)
    \bigg /
    \text{Var}\left(U_{k_2}^{(\text{wlr}_i)} \right)
  }.
\end{equation*}
Note that the above equation is asymptotically true only under the null hypothesis when the independent increment property 
$
  \text{Cov} 
  \left(
  U_{k_1}^{(\text{wlr}_i)}, U_{k_2}^{(\text{wlr}_i)} \right) 
  = 
  \text{Var} \left( U_{k_1}^{(\text{wlr}_i)} \right)
$ 
is asymptotically true \citep{tsiatis1981asymptotic}.
Under the alternate hypotheses, although the independent increment property is not strictly satisfied, we find the above equation almost numerically holds under the local alternative condition or when the events are not too frequent \citep{wang2021simulation}.

The third type of correlation is correlation between different analyses and different tests. For two analyses $1 \leq k_1 \leq k_2 \leq K $ and two WLR test with the weight of FH$(p_i, q_i)$ and FH$(p_j, q_j)$, as shown in \cite{wang2021simulation} and \cite{ghosh2022robust}, the correlation 
is 
\begin{eqnarray*} 
  \nonumber
  \text{Corr}
  \left(
    Z_{k_1}^{\left(\text{wlr}_i \right)},
    Z_{k_2}^{\left(\text{wlr}_j \right)}
  \right) 
  =
  \text{Corr}
  \left(
    Z_{k_1}^{\left(\text{wlr}_i \right)},
    U_{k_1}^{\left(\text{wlr}_j \right)}
  \right)
  \text{Corr}
  \left(
    Z_{k_1}^{\left(\text{wlr}_j \right)},
    Z_{k_2}^{\left(\text{wlr}_j
    \right)}
  \right).
\end{eqnarray*}
With the above three correlations, one can get asymptotic distribution of 
$
  \{ Z_k^{(\text{wlr}_i} \}_{k=1, \ldots, K; \; i = 1, \ldots, L}
$
by using either distribution-based prediction or data-driven estimation, The obtained outcomes can be utilized to compute the boundaries and probabilities of crossing. Comprehensive illustrations of these calculations will be discussed in Section \ref{sec: design char}.

\section{Group sequential design}
\label{sec: design char}

In this section, we discuss the derivation of design characteristics with the three tests introduced in Section \ref{sec: method}.
Design characteristics include spending functions and boundary calculations, type I error, power, sample size, and the number of events.

\subsection{Boundaries}
\label{sec: design char -- bound}

In group sequential designs, there are two sets of boundaries: upper boundaries and lower boundaries. The upper boundaries are referred to as \textit{efficacy boundaries} and the lower boundaries are called \textit{futility boundaries}. To select these boundaries, there are commonly two options: (i) pre-fixed boundaries; (ii) boundaries derived from the spending functions.
If there is a need to modify boundaries based on evolving information during analyses, it is advisable to avoid using the option (i) and opt for the option (ii) instead. Within this section, we will examine the calculation of boundaries using three tests that were previously introduced in Section \ref{sec: method}. Specifically, our attention will be directed toward situations in which boundaries are determined through the utilization of spending functions.  

The upper boundary $\mathbf{b} = (b_1, b_2, \ldots, b_K)^\top$ is decided by the type I error.
To spend the type I error $\alpha$ with the $K$ analyses, the monotone increasing error spending function $\alpha(t)$ with $t \geq 0$ is used, where $\alpha(0) = 0$ and $\alpha(t) = \alpha$ for any $t \geq 1$.
Without loss of generality, we derive the upper boundary with non-binding futility bound that is commonly used in practice. In other words, the lower bound is negative infinity under the null hypothesis. 
Specifically, the boundary at the $k$-th look can be calculated by solving
$
  b_k 
  = \left\{
  b_k :
  \text{Pr}
  \left(
    \mathcal Z_k \geq b_k, \;
    \cap_{i=1}^{k-1}
    \mathcal Z_i < b_i
    \;|\;
    H_0
  \right)
  =
  \alpha(t_k) - \alpha(t_{k-1})
  \right\},
$
where $t_k$ is the information fraction at the $k$-th look and $\mathcal Z_k \in \{Z_k^{(lr)}, Z_k^{(wlr)}, Z_k^{(mc)}\}$ is the test statistics depending on the selected test.

The lower boundary $\mathbf{a} = (a_1, a_2, \ldots, a_K)^\top$ is often determined by type II error $\beta$.
To spend $\beta$ with $K$ analyses, a monotone increasing error spending function $\beta(t)$ is used. The boundary at the $k$-th analysis is
$
  a_k
  =
  \left\{
  a_k
  :
  \text{Pr}
  \left(
    \{ \mathcal Z_k \leq a_k \}, \;
    \cap_{i=1}^{k-1}
    \{a_i < \mathcal Z_i < b_i\}
    \; | \;
    H_1
  \right)
  =
  \beta(t_k) - \beta(t_{k-1})
  \right\}.
$

For both the lower and upper boundary, with the asymptotic distribution of $\{\mathcal Z_k\}_{k = 1, \ldots, K}$ in Section \ref{sec: method}, it is feasible to resolve $\mathbf b$ and $\mathbf a$. In Appendix \ref{appendix: examples to solve upper and lower bounds}, we provide couple of examples.

\subsection{Type I error and power}
\label{sec: design char -- cross prob}

With the known bounds from Section \ref{sec: design char -- bound}, we can further derive the boundary crossing probabilities.
For example, the type I error and power at the final analysis is
\begin{eqnarray*}
  \text{type I error}
  & = &
  \sum_{k=1}^{K} 
  \text{Pr}
  \left(
    \{\mathcal Z_k \ge b_k\}, \;
    \cap_{j=1}^{k-1} \{\mathcal Z_j \le b_j\} \; | \;H_0 
  \right);\\
  \text{power}
  & = &
  \sum_{k=1}^{K}
  \text{Pr}
  \left(
    \{\mathcal Z_k > b_k\},\; 
    \cap_{j=1}^{k-1} \{a_j \le \mathcal Z_j \le b_j\} 
    \; | \;
    H_1
  \right).
\end{eqnarray*}
where $\mathcal Z_k \in \{Z_k^{\text{(lr)}}, Z_k^{\text{(wlr)}}, Z_k^{\text{(mc)}}\}$ is the test statistic depending on the selected test. 
To solve the above crossing probability explicitly, we utilize the distribution of $\mathcal Z_k$ in Section \ref{sec: method}. Examples in Appendix \ref{appendix: examples to solve upper and lower bounds} illustrate detailed calculation of power and type I error in practice.

\subsection{Sample Size and number of events}

\label{sec: design char -- sample size event}
In this section, we discuss the sample size and the number of events within a fixed study duration $\tau$. When considering a fixed study duration $\tau$, there are generally two approaches: we refer to as the \textit{d-n method} and the \textit{n-d method}. The d-n method involves initially estimating the expected number of events and subsequently enrolling subjects until this expected number of events is reached. This approach was employed by \cite{LachinFoulkes}. The n-d method follows a different logic, as it calculates the sample size first and then determines number of events by multiplying sample size by the failure probability. Essentially, the failure probability serves to estimate expected events. 

The LR test uses the d-n method to first calculate the number of events as
$$
    d 
    = 
    \sum_{m=1}^{M} 
    E \left( \bar{n}(\tau_{m-1}, \tau_m) \right), 
$$
where
$
  E \left( \bar{n}(\tau_{m-1}, \tau_m) \right)
  =
  G_{M+1-m}
  d_m
  +
  \frac{\lambda_m Q_{m-1} \gamma_{M+1-m} }{\lambda_m + \eta_m}
  \left(
    \tau_m -\tau_{m-1}
    -
    \frac{1 - q_m}{\lambda_m + \eta_m}
  \right)
$
is the expected number of events in the interval $[\tau_{m-1}, \tau_m)$. Here $q_m, Q_m, d_m$ are recursively defined as
$
    q_m = e^{-(\lambda+\eta_m)(\tau_m - \tau_{m-1})}
$, 
$
    Q_m = \prod_{j=1}^m q_j
$, 
and 
$
  d_m = \frac{\lambda_m Q_{m-1}}{\lambda_m + \eta_m}\left( 1 - e^{-(\lambda_m + \eta_m)(\tau_m - \tau_{m-1})}\right).
$
The detailed derivation of the above formulation is shown in Appendix \ref{proof: ahr -- sample size and events}. And the sample size is the one that achieves the above-expected number of events.

Both WLR test and MaxCombo tests use the n-d method to first calculate sample size as
  $$
    N
    =
    \inf
    \left\{
      N:
      1 - \beta
      =
      \text{Pr}\left(
      Z_1 \ge b_1 
      \right)
      +
      \sum_{k=2}^{K} 
      \text{Pr}\left(
      \cap_{j=1}^{k-1} a_j < Z_j < b_j,
      Z_k \ge b_k \; 
      \right)
    \right\}.
  $$
With the sample size available, the number of events is
  $
    d = n v(\tau),
  $
  where $\tau = \tau_a + \tau_f$ is the duration total of the study.
  We note that $v(t)= p_0 v_0(t) + p_1 v_1(t)$ is the failure probability.
  Here $v_j(t)$ represents the probability that a subject in arm $j$ will experience an event by time $t$, i.e., $v_j(t) = \int_0^t f_j(s) ( 1 - G_j(s)) H(\min\{\tau_a, \tau - s\}) ds$.

\subsection{Average HR}

When the HR is not constant over time, average HR is a useful metric of treatment effect: it represents the average benefit over the period of observation \citep{leon2020weighted}.

\paragraph{\underline{Average HR derived from the AHR method.}} 
To calculate the average HR (denoted as $\varphi^{\text{(lr)}}$), we weight the assumed individual HR by the expected number of events (inverse variance) in the corresponding interval under design assumptions:
  $$
    \varphi^{\text{(lr)}}
    =
    \sum_{m=1}^M w_m \varphi_m,
  $$
  where 
  $
    \varphi_m 
    =
    \log(\lambda_{1,m} / \lambda_{0,m})
  $
  is the log hazard ratio in the $m$-th interval.
  In practice, one can estimate $\varphi^{\text{(lr)}}$ as 
  $ 
    \widehat\varphi^{\text{(lr)}} = \sum_{m=1}^M \widehat w_m \widehat \varphi_m,
  $
  where
  $
    \widehat\varphi_m
    = 
    \log(d_{1,m}/T_{1,m}) - \log(d_{0,m}/T_{0,m}) 
  $
  and 
  $
    \widehat w_m
    =
    \frac{1}{1/d_{0,m}+1/d_{1,m}}
    \bigg/
    \sum_{i=1}^M
    \frac{1}{1/d_{0,i}+1/d_{1,i}}.
  $
  Here $\widehat\varphi_m$ is estimated from the partial likelihood function (see Appendix \ref{appendix: proof -- ahr -- ahr asy dist}) with $d_{i, m}$ as the number of events in group $i$ in the $m$-th interval. Weight $\widehat w_m$ is  an inverse variance weights.  
  For this estimated $\widehat \varphi^{\text{(lr)}}$, we have
  $
    \widehat\varphi^{\text{(lr)}}
    \overset{\cdot}{\sim}
    \text{Normal}(\varphi^{\text{(lr)}}, \; \mathcal{I}^{-1}),
  $
  where $\mathcal{I} = \sum_{m = 1}^M \left( \frac{1}{d_{0,m}} + \frac{1}{d_{1,m}} \right)^{-1}$.
  Details to derive the above are in Appendix \ref{appendix: proof -- ahr -- ahr asy dist}.

  \paragraph{\underline{Average HR derived from the WLR test.}} In the context of the WLR test introduced in Section \ref{sec: test -- wlr}, the computation of the average HR (denoted as $\varphi^{(\text{wlr})}$) draws inspiration from $\Delta_k$ as presented in \eqref{eq:WlrDletaK}. It is important to note that the last term in this equation represents hazard difference $\lambda_1(s) - \lambda_0(s)$.
  To facilitate calculations, we employ  Taylor expansion to approximate this hazard difference as 
  $
  \log
  \left[ \lambda_1(s)/\lambda_0(s) \right],
  $
  which gives an approximated $\Delta_k$ as 
  $$
    \Delta_k 
    \approx
    \int_{0}^{\tau_k} w(s)
    \frac{
      p_{0,k} \pi_{0,k}(s) \;
      p_{1,k} \pi_{1,k}(s)
    }{\pi_k(s)^2}
    \log\left(\frac{\lambda_1(s)}{\lambda_0(s)} \right) v'(s) ds,
  $$
  see derivation at Appendix \ref{appendix: wlr's average HR}.
  The difference between the regular hazard ratio and the above $\Delta_k$ only lies in the coefficients 
  $
  w(s)
  \frac{p_{0}\pi_{0}(s)p_{1}\pi_{1}(s)}{\pi(s)}.
  $
  These coefficients weight the individual hazard ratio based on the at-risk probability. 
  Thus, normalizing these coefficients of $\Delta_k$ gives an approximated average HR:
  $$
    \varphi^{(\text{wlr})}
    = 
    \Delta_K
    \bigg/
    \int_0^{\tau}
    w(s)
    \frac{p_{0,K} \pi_{0,K}(s) \;
    p_{1,K} \pi_{1,K}(s)}{\pi_K(s)^2} v'(s)
    ds,
   $$
   where $\Delta_k$ is represented by \eqref{eq:WlrDletaK}, and $\tau$ is the total duration of the study. The term $p_{j,K} = n_{j,K}/(n_{0,K} + n_{1,k})$ denotes the randomization probability for group $j$. Furthermore, $\pi_K(t) = p_{0,K} \pi_{0,K}(t) + p_{1,K}\pi_{1,K}(t)$ defines the overall at-risk probability, where $\pi_{j,K}(t)$ represents the expected at-risk probability of group $j$. Details of the above statement are in Appendix \ref{appendix: wlr's average HR}.
   A bridge to connect average HR by AHR method and WLR test is provided in Appendix \ref{appendix: 2 beta bridge}.

\subsection{Information fraction}

For the $k$-th analysis, we denote the treatment effect as $\theta_k$. In the LR test using the AHR method, the statistical information for the estimate $\widehat{\theta_k}$ is given by:
$
  \mathcal I_k = 1 /  \text{Var}(\widehat\theta_k).
$
The $t_k$ is so-called information fraction at analysis $k$ in that $t_k = \mathcal I_k / \mathcal I_K$.

For the WLR test, the statistical information and information fraction have been discussed extensively in the literature. For instance, studies such as \cite{gillen2005information}, \cite{brummel2014flexibly}, \cite{kundu2020comments}, and \cite{kundu2021information} have presented the statistical information of WLR tests using the FH weights. In this paper, we focus on the statistical information of WLR tests with general weights, which is expressed as $n \sigma_k^2$ for the $k$-th analysis, where $\sigma_k^2$ is provided in equation \eqref{equ: sigma of wlr}.

Regarding the MaxCombo test, it combines several tests, each with its own information fraction. 
In our developed gsDesign2 R package, we consider the information fraction of the MaxCombo test for spending to be the minimal information fraction among the tests it combines.
However, the full correlation matrix for all tests at all analyses are used for the asymptotic normal distribution used to compute boundary crossing probabilities.

\section{Simulations}
\label{sec: simulation}

\begin{figure}
    \centering
    \includegraphics[width=0.7\linewidth]{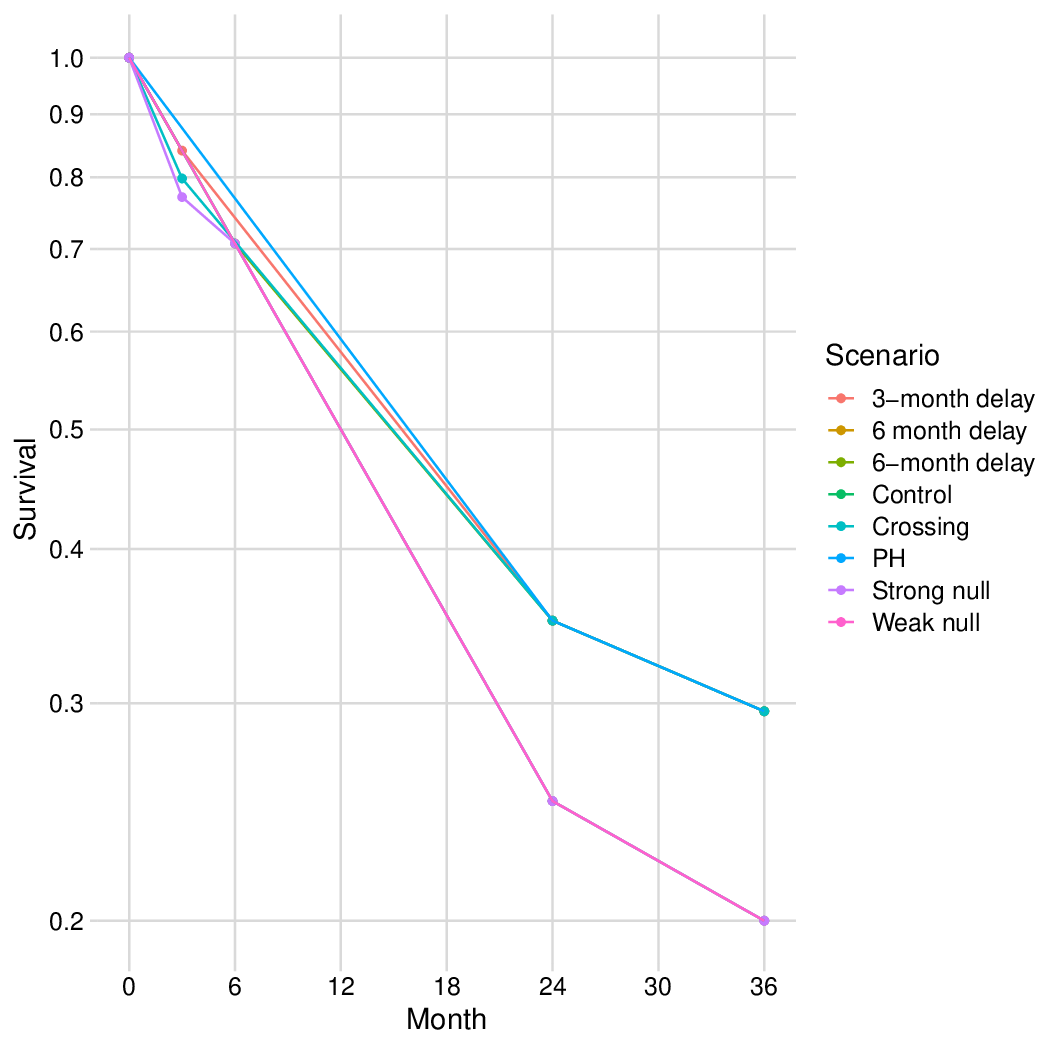}
    \caption{Survival of 6 different scenarios}
    \label{fig:sim -- scenario plot}
\end{figure}

We assume 6 different design assumption scenarios with a constant that there is an assumed underlying survival of 35\% in the experimental group compared to 25\% in the control group 2 years after start of treatment; lower event rates are assumed from 2 years through 3 years, but maintaining the same cumulative hazard ratio at 3 years. The 6 different scenarios are: proportional hazards, 3-month dela, 6-month delayed effect, hazard ratio of 1.3 for 3 months followed by a constant hazard ratio, the weak null hypothesis where experimental treatment outcomes have the same underlying distribution as control, and strong null hypothesis with a hazard ratio of 1.5 (experimental/control) for 3 months with a constant hazard ratio of 0.5 to equalize survival by 6 months in each treatment group, and with a hazard ratio of 1 thereafter. 
The survival curves of the aforementioned 6 scenarios are presented in Figure \ref{fig:sim -- scenario plot}.
Other assumptions for the scenario are an expected enrollment duration of 1 year and a total study duration of 3 years. The control group time-to-event distribution is assumed to be exponential with a median of 12 months in all scenarios. A constant exponential dropout rate of 0.001 is assumed for both treatment groups throughout.

For statistical testing, we consider 7 possible methods: (1) Logrank test; (2) Fleming-Harrington (weighted logrank) test with $\rho=0, \gamma=0.5$ (FH(0, 0.5)); (3) MaxCombo test with the logrank and FH(0,0.5) tests; (4) modestly weighted logrank test with $t^* = 12$ and maximal weight of 2 ($w_{\max} = 2$); (5) Weighted logrank with zero-early weighting for 3 months with weight 1 thereafter \citep{xu2017designing}; (6) RMST (restricted mean survival); and (7) Comparison of survival at 2-years (milestone test).

\begin{figure}
    \centering
    \includegraphics[width=0.6\linewidth]{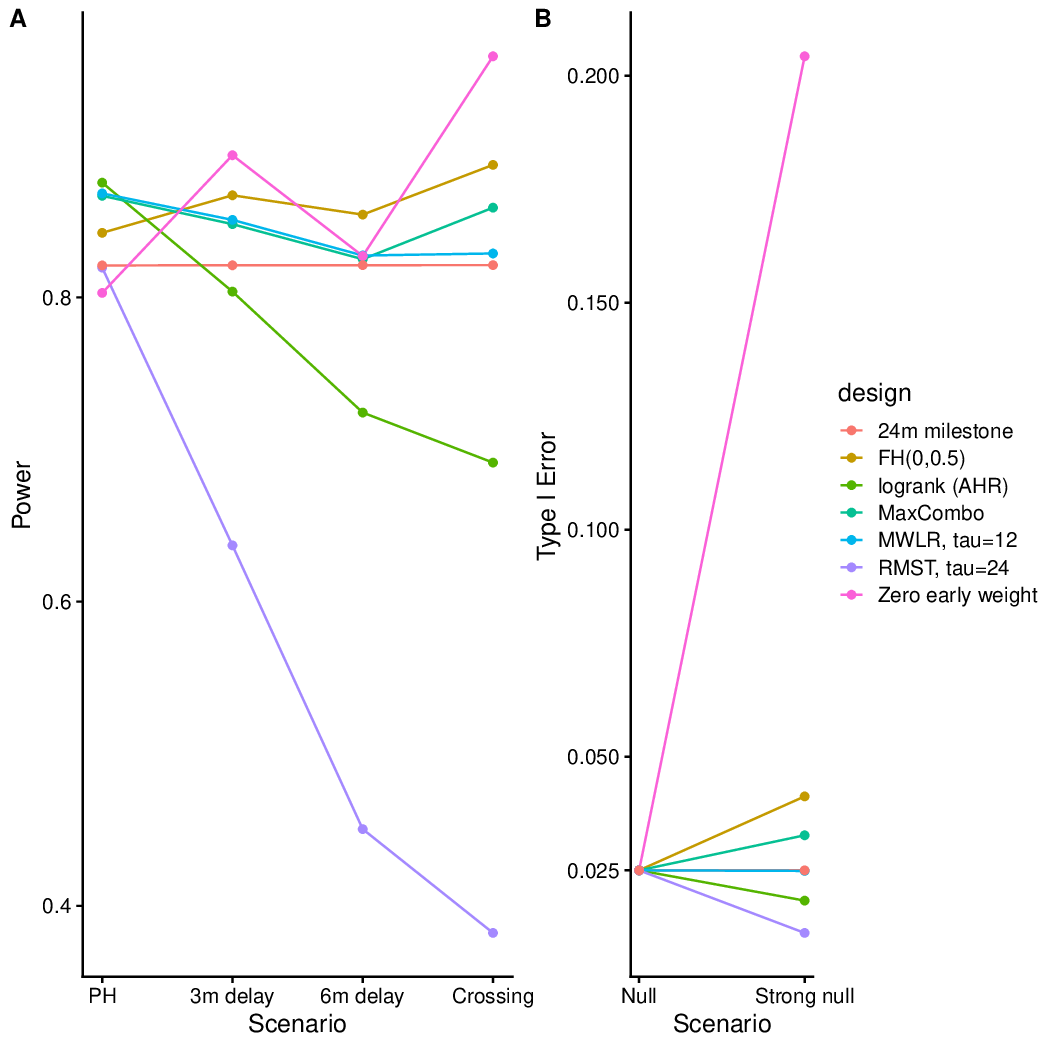}
    \caption{Asymptotic power and type I error of 6 discussed scenarios under the sample size of 698 and study duration of 3 years\label{fig: sim -- asy power and type I error}}
\end{figure}

We choose the sample size ($N= 698$) of the modestly weighted logrank test. We compare power across the different scenarios with common underlying benefit at 2 and 3 years. We also compute Type I error under the null and strong null hypotheses. The asymptotic power and type I error is provided in Table \ref{tab: sim -- asy power and type I error} and visualized in Figure \ref{fig: sim -- asy power and type I error}. The specific strong null hypothesis chosen here is comparable to an example of \cite{magirr2023strong}; it is chosen to demonstrate excess Type I error for weighted logrank tests with 0 early weights; this includes Fleming-Harrington \citep{FH2011} tests with $\rho=0$ such as the FH(0,0.5) studied here and the zero-early-weighting test proposed by \cite{xu2017designing}. The Type I error issue extends to MaxCombo tests \citep{roychoudhury2021robust}. As noted by \cite{Magirr2019}, for the modestly weighted logrank tests the Type I error issue does not exist. The zero-early-weighted tests could be justified such as when patients in a personalized cancer vaccine trial have identical treatment during an early vaccine manufacturing period.

\begin{table}[htbp]
\caption{Asymptotic power and type I error of 6 discussed scenarios under the sample size of 698 and study duration of 3 years\label{tab: sim -- asy power and type I error}}
\begin{adjustbox}{max width=1\textwidth}
\begin{threeparttable}
\begin{tabular}{c|cccccccc }
        \hline
        Scenario
        & Logrank
        & FH(0,0.5)
        & MaxCombo \textsuperscript{1}
        & MWLR(12) \textsuperscript{2}
        & Zero early weight	\textsuperscript{3}
        & RMST(24) \textsuperscript{4}	
        & Milestone (24) \textsuperscript{5}	 \\
        \hline
        PH
        & 0.875	&0.842	&0.867	&0.868	&0.803	&0.820	&0.821 \\
        3-month delay
        &0.804	&0.867	&0.848	&0.851	&0.893	&0.637	&0.821 \\
        6 month delay	
        &0.724	&0.854	&0.825	&0.828	&0.827	&0.450	&0.821 \\
        Crossing	
        &0.691	&0.887	&0.859	&0.829	&0.958	&0.382	&0.821 \\
        Weak null	
        &0.025	&0.025	&0.025	&0.025	&0.025	&0.025	&0.025 \\
        Strong null	
        &0.018	&0.041	&0.033	&0.025	&0.204	&0.011	&0.025 \\
        \hline
\end{tabular}
\begin{tablenotes}
    \item[1] MaxCombo test with the logrank and FH(0,0.5) tests.
    \item[2] Modestly weighted logrank test with $\tau = 12$.
    \item[3] Weighted logrank with zero-early weighting for 3 months with weight 1 thereafter.
    \item[4] RMST difference at month 24.
    \item[5] Comparison of survival at 2-years (milestone test).
   \end{tablenotes}
  \end{threeparttable}
\end{adjustbox}
\end{table}

To verify the above asymptotic results, we run 1 million simulations for numerical verification per test per scenario, where the simulation results are summarized in Table \ref{tab: sim -- simulated power and type I error}. We find the simulations verify asymptotic calculations.
We consider the weak and strong null hypotheses for the MaxCombo test that tests with the maximum of the logrank and FH(0, 0.5) tests. Type I error by the asymptotic calculations above was inflated under the strong null hypotheses.

\begin{table}[htbp]
\caption{Numerical power and type I error of 6 discussed scenarios under the sample size of 698 and study duration of 3 years\label{tab: sim -- simulated power and type I error}}
\begin{adjustbox}{max width=1\textwidth}
\begin{threeparttable}
\begin{tabular}{c|cccccccc }
        \hline
        Scenario
        & Logrank
        & FH(0,0.5)
        & MaxCombo \textsuperscript{1}
        & MWLR(12) \textsuperscript{2}
        & Zero early weight	\textsuperscript{3}
        & RMST(24) \textsuperscript{4}	
        & Milestone (24) \textsuperscript{5}	 \\
        \hline
        PH
        & 0.876 & 0.836 & 0.866 & 0.863 & 0.798 & 0.818 & 0.820\\
        3-month delay
        & 0.803 & 0.862 & 0.848 & 0.846 & 0.891 & 0.635 & 0.820\\
        6 month delay	
        & 0.722 & 0.849 & 0.825 & 0.823 & 0.823 & 0.450 & 0.820\\
        Crossing	
        & 0.686 & 0.883 & 0.858 & 0.823 & 0.958 & 0.382 & 0.820\\
        Weak null	
        & 0.025 & 0.025 & 0.025 &0.025 & 0.025 & 0.025 & 0.025\\
        Strong null	
        & 0.016 & 0.042 & 0.033 &0.025 & 0.205 & 0.011 & 0.025\\
        \hline
\end{tabular}
\begin{tablenotes}
    \item[1] MaxCombo test with the logrank and FH(0,0.5) tests.
    \item[2] Modestly weighted logrank test with $\tau = 12$.
    \item[3] Weighted logrank with zero-early weighting for 3 months with weight 1 thereafter.
    \item[4] RMST difference at month 24.
    \item[5] Comparison of survival at 2-years (milestone test).
   \end{tablenotes}
  \end{threeparttable}
\end{adjustbox}
\end{table}

\section{Examples}
\label{sec: application}

The example discussed in this section has a 12-month enrollment period with a monthly enrollment rate of 500/12. Our study aims to achieve a targeted power of 90\% while maintaining a controlled type I error rate of 0.025. Additionally, we consider the presence of a delayed treatment effect, characterized by an HR of 1 for the first 4 months and 0.6 thereafter. The control arm has a median survival of 15 months, and the dropout rate remains consistent at 0.001 across all study arms throughout the duration of the study. The one-sided group sequential design discussed comprising 4 analyses conducted at the 12, 20, 28, and 36 months. 

\begin{figure}[htbp]
  \begin{center}
    \includegraphics[width = 0.6\textwidth]{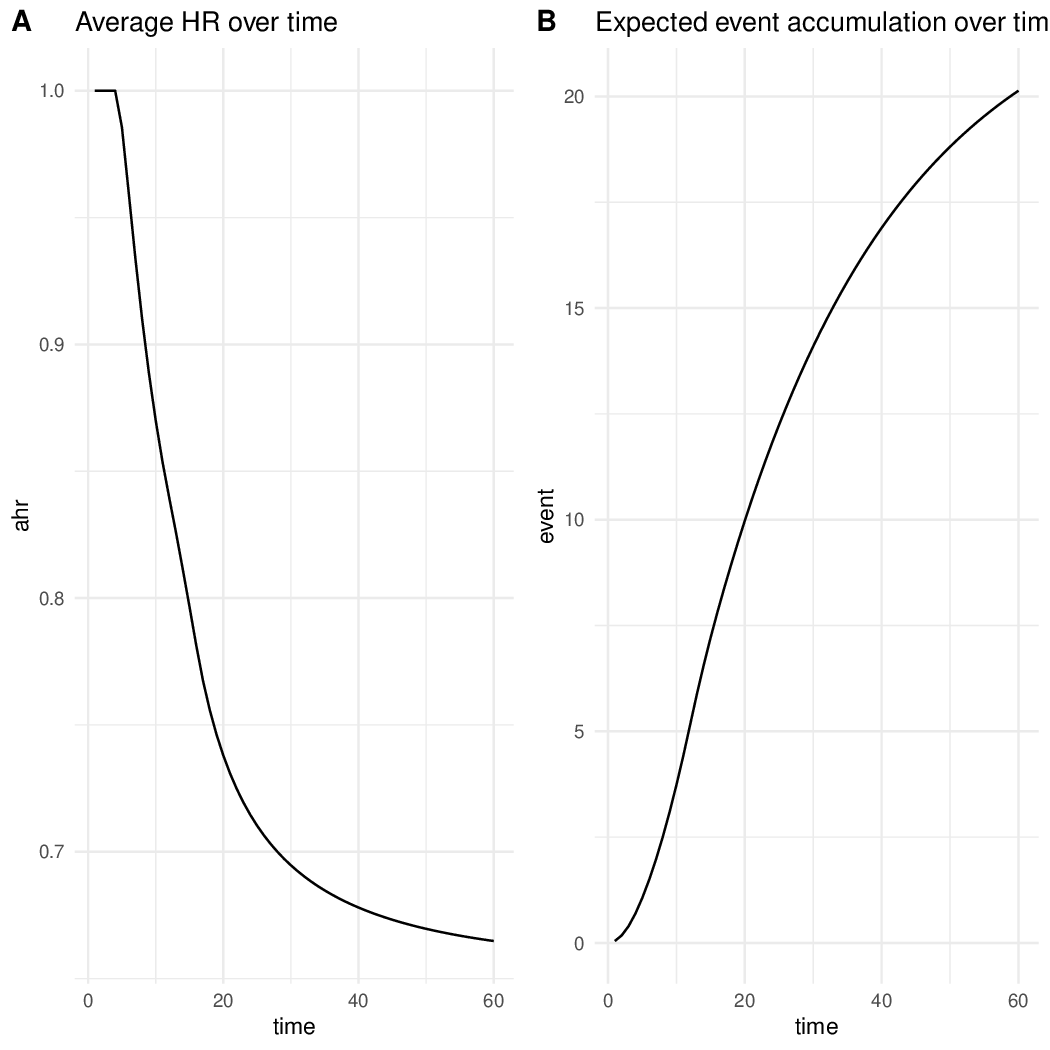} \\
    \caption{Average HR as a function of study duration (left) and expected event accumulation as a function of study duration (right) \label{fig: example_ahr_visualization}}
  \end{center}
\end{figure}

To gain a better understanding of the example described above, it is helpful to start with visualization. On the left-hand side of Figure \ref{fig: example_ahr_visualization}, we present a plot showing the AHR as a function of trial duration, taking into account the modified enrollment required to achieve the desired power for the trial. Here, we observe an AHR of 1 in the first few months, reflecting an assumed delayed treatment effect within the initial 4 months. On the right-hand side of Figure \ref{fig: example_ahr_visualization}, we display the expected event accrual over time. Both plots offer valuable insights: a key design consideration involves selecting the trial duration based on factors such as the extent of AHR improvement over time versus the urgency of completing the trial as quickly as possible. It is important to note that longer follow-up duration can lead to a decrease in the required sample size.

For the first 3 analyses, the regular logrank test is implemented. At the final analysis, we use MaxCombo test with the logrank and FH(0,0.5) tests. The asymptotic design to get a 90\% power is summarized in Table \ref{tab: example -- design summary}.

\begin{table}[htbp]
\caption{Bound summary for MaxCombo \textsuperscript{1} design \label{tab: example -- design summary}}
\begin{adjustbox}{max width=0.98\textwidth}
\begin{threeparttable}
\begin{tabular}{ccccccccc }
    \hline 
    \multirow{2}{*}{Bound} 
    & \multirow{2}{*}{Z}  
    & \multirow{2}{*}{Nominal p\textsuperscript{2}}  
    & \multicolumn{2}{c}{Cumulative boundary crossing probability} \\
    & & & Alternative hypothesis & Null hypothesis \\
    \hline 
    \rowcolor{gray}
    \multicolumn{5}{l}{\textcolor{white}{Analysis: 1 Time: 12 N: 643.5 Event: 138.2 AHR\textsuperscript{3}: 0.84 Event fraction \textsuperscript{4}: 0.32}} \\
    \hline
    Efficacy & 6.18 & 0.0000 & 0.0000 & 0.0000 \\
    \hline 
    \rowcolor{gray}\multicolumn{5}{l}{\textcolor{white}{Analysis: 2 Time: 20 N: 643.5 Event: 267.6 AHR: 0.74 Event fraction: 0.63}} \\
    \hline 
    Efficacy &	3.37 & 0.0004 & 0.1805 & 0.0004 \\
    \hline 
    \rowcolor{gray}\multicolumn{5}{l}{\textcolor{white}{Analysis: 3 Time: 28 N: 643.5 Event: 359.2 AHR: 0.7 Event fraction: 0.84}} \\
    \hline 
    Efficacy & 2.42	 & 0.0077 & 0.8240 & 0.0077 \\
    \hline 
    \rowcolor{gray}\multicolumn{5}{l}{\textcolor{white}{Analysis: 4 Time: 36 N: 643.5 Event: 426.4 AHR: 0.68 Event fraction: 1}} \\
    \hline 
    Efficacy & 2.02	 & 0.0219 & 0.9900 & 0.0250 \\
    \hline
    \end{tabular}
    \begin{tablenotes}
    \footnotesize
    \item[1] For the 3 interim analyses, the logrank test is utilized. For the final analysis, the MaxCombo test combining the logrank and FH(0,0.5) tests is used.
    \item[2] One-sided p-value for experimental vs control treatment. Value $<$ 0.5 favors experimental, $>$ 0.5 favors control.
    \item[3] AHR is under regular weighted log rank test.
    \item[4] The minimal information fraction of logrank test of  FH(0, 0.5) test is used to decided the alpha spending.
   \end{tablenotes}
  \end{threeparttable}
\end{adjustbox}
\end{table}


\section{Discussion}
\label{sec: discussion}

Group sequential design has been widely used in clinical trials, particularly for time-to-event endpoints. Recent results from immunotherapy-based oncology trials have highlighted the presence of NPH. Thoroughly assessing NPH scenarios during the trial design stage becomes paramount to appropriately power trials.
This paper undertakes a comprehensive exploration of three commonly employed NPH methods for group sequential design. These methodologies have been seamlessly integrated into the \cite{gsDesign2} and are futher supported by the simulation capabilities of the \cite{simtrial}. Furthermore, our ongoing efforts involve expanding the functionality of the R package to design stratified clinical trials under NPH. 

\section*{Appendix}
\appendix

\section{A brief introduction of group sequential design}
\label{appendix: into of gsd}
One of the key challenges in group sequential design is the determination of the test boundary. For example, if we consider the upper bound $\{b_k\}_{k = 1, \ldots, K}$ that can strongly control the overall Type I error, say $\alpha = 0.05$.  
Using the alpha spending function approach \citep{demets1994interim} to specify how much Type I error $\alpha_k$ spent on the $k$ -th analysis with 
$$\sum_{k=1}^K \alpha_k= \alpha =0.05,$$
we can derive the
upper boundaries $\{b_k\}_{k = 1, \ldots, K}$ by sequentially solving 
$$
  \alpha_1 
  = 
  \text{Pr}(Z_k \ge b_k |H_0)
$$
and
$$
  \alpha_k = \text{Pr}(\{Z_k \ge b_k\}
  \cap_{j=1}^{k-1}\{Z_k < b_j\}|H_0) 
$$
for $2 \leq k \leq K$. 
Given the upper bound $\{b_k\}_{k=1, \ldots, K}$, the overall type I error is  
$$
  \sum_{k=1}^{K} 
  \text{Pr}(\{Z_k \ge b_k\}  \cap_{j=1}^{k-1}\{Z_k < b_j\}|H_0).
$$
To calculate the above probability, we need to know the joint distribution of these test statistics $\{Z_k\}_{k=1,\ldots, K}$.

Asymptotically, the sequence of test statistics $\{Z_k\}_{k=1,\ldots, K}$ 
is a normal process with independent increments \citep{scharfstein1997semiparametric} for many commonly used test statistics for continuous, binary, and survival outcomes. 
The joint distribution of $\{Z_k\}_{k=1,\ldots, K}$ is multivariate normal
with $E(Z_k)=\theta\sqrt{\mathcal{I}_k}$ and
$\text{Cov}(Z_i, Z_j)=\sqrt{\mathcal{I}_i / \mathcal{I}_j}$, $1 \le i \le j \le K$
with information level $\{\mathcal{I}_k\}_{k=1,\ldots, K}$ for the parameter $\theta$.
This specific multivariate normal distribution is called the \textit{canonical joint distribution} in Chapter 3 of \cite{JTBook}. 
With the canonical joint distribution, design characteristics can be derived in a unified approach analytically.

For the logrank test and weighted logrank test, the canonical joint distribution can be derived. However, the joint distribution is not multivariate normal for the MaxCombo test \citep{wang2021simulation}. 

\section{Deriving the asymptotic distribution of the log HR under the piecewise model in Section \ref{sec: test -- ahr}}
\label{appendix: proof -- ahr -- ahr asy dist}

We consider a piecewise parametric model, possibly with stratification.
We begin with parametric modeling for individual treatment groups and intervals, and then extend to estimation and testing across intervals. The methods are assumed to extend to Cox regression and logrank tests as summarized in the body of the paper.
For an individual interval $t_m$ and stratum $j$, and treatment $i$, the likelihood of $\lambda_{i,j,m}$ is
$$
  L(\lambda_{i,j,m})
  =
  \exp(-\lambda_{i,j,m}T_{i,j,m})\;
  \lambda_{i,j,m}^{d_{i,j,m}},
$$
where for stratum $j$, $d_{i,j,m}$ is the observed number of events for treatment group $\forall i \in \{0,1\}$ in $(t_{m-1}, t_m]$ and $T_{i,j,m}$ is the follow-up time (total time on test) in $(t_{m-1}, t_m]$.
The above likelihood function of $\lambda_{i,j,m}$ can be re-written as the likelihood function for $\gamma_{i, j, m} = \log(\lambda_{i,j,m})$, i.e.,
$$
  L(\gamma_{i,j,m})
  =
  \exp(-e^{\gamma_{i,m}}T_{i,j,m})\;
  e^{\gamma_{i,j,m}d_{i,j,m}}.
$$
This leads to a log-likelihood:
$$
  \ell(\gamma_{i,j,m})
  \triangleq
  \log\left( L(\gamma_{i,j,m}) \right)
  =
  -e^{\gamma_{i,j,m}}T_{i,j,m}
  +
  \gamma_{i,j,m}d_{i,j,m}.
$$
By setting the first derivative with respect to $\gamma_{i,j,m}$ 
$$
  \frac{\partial}{\partial \gamma_{i,j,m}}
  \ell(\gamma_{i,j,m})
  =
  -e^{\gamma_{i,j,m}}T_{i,j,m} + d_{i,j,m}
$$
to zero, we get the maximum likelihood estimate
\begin{equation}
  \hat\gamma_{i,j,m}
  =
  \log(d_{i,j,m}/T_{i,j,m}).
  \label{eq:AhrRateEstimation}
\end{equation}
The Fisher information for $\hat\gamma_{i,j,m}$ is 

\begin{equation}
\mathcal{I}(\widehat\gamma_{i,j,m})=-\hbox{E}
\left[
  \frac{\partial^2}{\partial \gamma_{i,j,m}^2} \ell(\hat\gamma_{i,j,m})
| \gamma_{i,j,m})
\right] 
= 
\hbox{E}(e^{\widehat\gamma_{i,j,m}}T_{i,j,m}| \gamma_{i,j,m}) 
= 
\hbox{E}(d_{i,j,m}| \gamma_{i,j,m}).
\end{equation}

Thus, the asymptotic variance of $\hat\gamma_{i,j,m}$ is
\begin{equation}
\hbox{Var}(\hat\gamma_{i,j,m})\doteq  1/\mathcal{I}(\widehat\gamma_{i,j,m}) = 1/\hbox{E}(d_{i,j,m}|\gamma_{i,j,m}).
\end{equation}

The calculation of E($d_{i,j,m}|\gamma_{i,j,m})$ is provided in detail at 
\url{https://merck.github.io/gsDesign2/articles/story-compute-expected-events.html}.

The asymptotic distribution of $\widehat\lambda_{i,j,m}$ is thus
\begin{equation}
  \log(\widehat\lambda_{i,j,m})
  \overset{\cdot}{\sim}
  \text{Normal}
  \left(
    \log(\lambda_{i,j,m}), \; 1/\hbox{E}(d_{i,j,m}|\lambda_{i,j,m})
  \right),
  \;\;
  \forall i \in \{0, 1\}.
  \label{eq:AhrIndividualHazardRateDist}
\end{equation}

We can estimate 
\begin{equation*}
  \widehat\varphi_{j,m}
  =
  \log\left( \frac{\widehat\lambda_{1,j,m}}{\widehat\lambda_{0,j,m}} \right)
  =
  \log(\hat\lambda_{1,j,m}) - \log(\hat\lambda_{0,j,m})
\end{equation*}
which leads to the asymptotic distribution of $\hat\varphi_{j,m}$:
\begin{equation*}
  \label{eq:AhrSignleLogHazardRatio}
  \widehat\varphi_{j,m}
  \overset{\cdot}{\sim}
  \text{Normal}
  \left(
    \varphi_{j,m},
    \frac{1}{\hbox{E}(d_{0,j,m}|\gamma_{i,j,m})} + \frac{1}{\hbox{E}(d_{1,j,m}|\gamma_{i,j,m})}
  \right)
  \;\; \forall m = 1,\ldots, M, j=1,\ldots,J.
\end{equation*}

\noindent
The Fisher information for $\hat\varphi_{j,m}$ is thus
\begin{equation*}
  \mathcal{I}(\hat\varphi_{j,m}) 
  = 
  \frac{\hbox{E}(d_{0,j,m}) \times \hbox{E}(d_{1,j,m})}{\hbox{E}(d_{0,j,m}) + \hbox{E}(d_{1,j,m})}.
\end{equation*}
Under the null hypothesis when where is no treatment effect, we have $\hbox{E}(d_{0,j,m}) = \hbox{E}(d_{1,j,m})$, which leads to $\mathcal{I}(\hat\varphi_{j,m}) = \hbox{E}(d_{1,j,m}) / 2 = \hbox{E}(d_{0,j,m}) / 2$. Under the alternative hypothesis, we have $\mathcal{I}(\hat\varphi_{j,m}) = \left[ 1/\hbox{E}(d_{1,j,m}) + 1/\hbox{E}(d_{0,j,m}) \right]^{-1}$.
The information weight for $\hat\varphi_{j,m}$ is
\begin{equation*}
  w_{j,m} 
  = 
  \mathcal{I}(\hat\varphi_{j,m})/\sum_{\ell=1}^M\sum_{a=1}^J
\mathcal{I}(\hat\varphi_{a,\ell}).
\end{equation*}
The above $w_{j,m}$ is utilized to build the average hazard ratio, i.e.,
\begin{equation*}
  \widehat\varphi^{(lr)}
  =
  \sum_{j=1}^J\sum_{m=1}^M
  w_{j,m} \widehat\varphi_{j,m}.
\end{equation*}
It is straightforward to show that the total information across strata (inverse of the variance of $\widehat\varphi^{(lr)}$) is
\begin{equation*}
  \mathcal{I}(\widehat\varphi^{(lr)}) 
  = 
  \sum_{j=1}^J\sum_{m=1}^M
  \mathcal{I}(\widehat\varphi_{j,m}).
\end{equation*}


\section{Examples of solving upper and lower spending bounds and crossing probabilities}
\label{appendix: examples to solve upper and lower bounds}

  In the context of a group sequential design involving $K$ analyses and utilizing error spending functions $\alpha(t)$ and $\beta(t)$ for controlling type I and type II errors, respectively, the AHR method provides upper bound  and lower bound  estimates at the initial analysis as
  $$ 
  \begin{array}{ccl}
    b_1 
    =
    \left\{
    b:
    \alpha(t_1) = 
    \text{Pr}
    \left(
      Z_1^{(\text{lr})} \geq b
      \; | \;
      H_0
    \right)
    \right\} \\
    a_1 
    =
    \left\{
    a:
    \beta(t_1)
    = 
    \text{Pr}
    \left(
      Z_1^{(\text{lr})} \leq a
      \; | \;
      H_1
    \right)
    \right\}
  \end{array}
  $$
  Since $Z_1^{(\text{lr})}$ follows a normal distribution as shown in Section \ref{sec: test -- ahr}, the above equation can be directly solved by integration. Generally, for the $k$-th analysis ($k = 2, \ldots, K$), the upper and lower bounds can be solved as
  \begin{equation*}
  \begin{array}{lllll}
    b_k
    = 
    \left\{
    b:
    \alpha(t_k) - \alpha(t_{k-1}) 
    =
    \text{Pr}
    \left(
      \underset{1\leq k' < k}{\cap}
      Z_{k'}^{(\text{lr})} < b_{k'}, \; Z_k^{(\text{lr})} \geq b
      \; | \;
      H_0
    \right) 
    \right\} \\
    a_k
    = 
    \left\{
    a:
    \beta(t_k) - \beta(t_{k-1}) 
    =
    \text{Pr}
    \left(
      \underset{1\leq k' < k}{\cap}
      Z_{k'}^{(\text{lr})} > a_{k'}, \; Z_k^{(\text{lr})} \leq a
      \; | \;
      H_1
    \right) 
    \right\}.
  \end{array} 
  \end{equation*}
  The same reasoning can be extended to the WLR test by substituting $Z_k^{(\text{lr})}$ with $Z_k^{(\text{wlr})}$ and employing the asymptotic distribution described in Section \ref{sec: test -- wlr}.

  Considering $Z_k^{(\text{mc})}$ involves a maximum operator, there is no canonical joint multivariate normal distribution. However, the asymptotic normal distribution of $Z_k^{(\text{wlr}_i)}$ can be used to derive the type I error, power, and other design characteristics (see example below).
\begin{example}
\label{example: design char -- bound - 1st analysis}
  In the context of a group sequential design involving $K$ analyses and utilizing error spending functions $\alpha(t)$ and $\beta(t)$ for controlling type I and type II errors, respectively, the MaxCombo test (consisting of $L$ WLR tests with FH weights) provides the upper bound $b_1$ at the first analysis as
  $$
    b_1
    = 
    \left\{
    b:\;
    \alpha(t_1) = 
    \text{Pr}
    \left(
      Z_1^{(\text{mc})} \geq b
      \; | \;
      H_0
    \right)
    \right\}.
  $$
  Since $Z_1^{(\text{mc})} = \max\left\{ Z_1^{(\text{wlr}_1)}, \ldots, Z_1^{(\text{wlr}_L)} \right\}$, we can further simplify the above equation as
  $$
    b_1
    =
    \left\{
    b: 
    \alpha(t_1)  
    =
    1
    -
    \text{Pr}
    \left(
      \cap_{\ell = 1}^L
      Z_1^{(\text{wlr}_\ell)}
      <
      b
      \; | \;
      H_0
    \right)
    \right\}.
  $$
  Given the known distribution of $\{Z_1^{(\text{wlr}_\ell)}\}_{\ell = 1, \ldots, L}$ (as described in Section \ref{sec: test -- mc}), the above equation can be solved using multiple integration techniques. Likewise, the value of the lower bound $a_1$ can be obtained by solving for $a_1$ solving 
  \begin{eqnarray*}
    a_1 = 
    \left\{
    a:
    \beta(t_1)
    = 
    \text{Pr}
    \left(
      Z_1^{(\text{mc})} \leq a
      \; | \;
      H_1
    \right) 
    \right\}
    = 
    \left\{
    \text{Pr}
    \left(
      \cap_{\ell=1}^L
      Z_1^{(\text{wlr}_\ell)} \leq a
    \; | \;
    H_1
    \right)
    \right\}.
  \end{eqnarray*}
  With the known values for $a_1$ and $b_1$, we can further calculate $a_2$ and $b_2$, which is presented in Example \ref{example: maxcombo solving a2 b2}.
\end{example}

\begin{example}
\label{example: maxcombo solving a2 b2}
  In accordance with Example \ref{example: design char -- bound - 1st analysis}, assuming known values for $a_1$ and $b_1$, when $k = 2$, the upper boundary $b_2$ can be determined by solving for $b_2$ by
  \begin{equation*}
  \begin{array}{lllll}
    b_2 
    & = &
    \left\{
    b: \;
    \alpha(t_2) - \alpha(t_1) 
    = 
    \text{Pr}
    \left(
      Z_1^{(\text{mc})} < b_1, Z_2^{(\text{mc})} \geq b_2
      \; | \;
      H_0
    \right) 
    \right\} \\
    & = &
    \left\{
    b: \;
    \alpha(t_2) - \alpha(t_1)  
    =
    \text{Pr}
    \left(
      Z_1^{(\text{mc})} < b_1
      \; | \;
      H_0
    \right)
    - 
    \text{Pr}
    \left(
      \cap_{\ell = 1}^L
      Z_1^{(\text{wlr}_{\ell})}
      < b_1,
      \;
      \cap_{\ell = 1}^L
      Z_2^{(\text{wlr}_{\ell})}
      < b_2
      \; | \;
      H_0
    \right).
    \right\} \\
  \end{array}
  \end{equation*}
  The first term $\left(Z_1^{(\text{mc})} < b_1\; | \;H_0\right)$ can be solved following Example \ref{example: design char -- bound - 1st analysis}. The second term can be computed using the distribution of $\{Z_k^{(\text{wlr}_{\ell})}\}_{k = 1, 2, \ell = 1, \ldots, L}$ in Section \ref{sec: test -- mc}.
  Similarly, the lower bound $a_2$ can be derived by solving $a_2$ by
  \begin{eqnarray*}
    a_2 
    & = &
    \left\{
    a: \;
    \beta(t_2) - \beta(t_1)
     = 
    \text{Pr}
    \left(
      a_1 < Z_1^{(\text{mc})} < b_1,
      Z_2^{(\text{mc})} \leq a_2
      \; | \;
      H_1
    \right)
    \right\} \\
    & = &
    \left\{
    a: \;
    \beta(t_2) - \beta(t_1)
    =
    \text{Pr}
    \left(
      Z_1^{(\text{mc})} < b_1, 
      Z_2^{(\text{mc})} \leq a_2
      \; | \;
      H_1
    \right)
    -
    \text{Pr}
    \left(
      Z_1^{(\text{mc})} < a_1, Z_2^{(\text{mc})} < a_2
      \; | \;
      H_1
    \right)
    \right\}
  \end{eqnarray*}
  Note that the above two terms can be both solved by integrating the distribution of $\{Z_k^{(\text{wlr}_{\ell})}\}_{k = 1,2,\; \ell = 1, \ldots, L}$ in Section \ref{sec: test -- mc}. 
\end{example}

\section{Deriving the average HR for the WLR test}
\label{appendix: wlr's average HR}
For the second term in \eqref{eq:WlrDletaK}, it is essentially a Harmonic mean 
$$
  \frac{
    p_{0,k} \pi_{0,k}(s) \;
    p_{1,k} \pi_{1,k}(s)
  }{
    \pi(s)
  }
  =
  1 \bigg/ \left[\frac{1}{p_{0,k} \pi_{0,k}(s)} +  \frac{1}{p_{1,k} \pi_{1,k}(s)}\right],
$$
which is used to weigh the difference between hazards.

  From \eqref{eq:WlrDletaK}, we have 
  \begin{eqnarray}
    \Delta_k  
    & = & \nonumber
    \int_{0}^{\tau_k} w(s)
    \frac{
      p_{0,k} \pi_{0,k}(s) \;
      p_{1,k} \pi_{1,k}(s)
    }{\pi_k(s)}
    \left[\lambda_{1}(s)-\lambda_{0}(s)\right] ds\\
    & = &
    \label{equ: delta k simple}
    \int_{0}^{\tau_k} w(s)
    \frac{
      p_{0,k} \pi_{0,k}(s) \;
      p_{1,k} \pi_{1,k}(s)
    }{\pi_k(s)^2}
    \underbrace{
      \left[\lambda_{1}(s)-\lambda_{0}(s)\right]\pi_k(s)
    }_{\mathcal A}ds.
  \end{eqnarray}
  For $\mathcal A$, one has
  \begin{eqnarray*}
      \mathcal A
      & = &
      \left[\lambda_1(s) - \lambda_0(s)\right] [p_0 \pi_{0,k}(s) + p_1 \pi_{1,k}(s)] \\
      & = &
      \left[\lambda_1(s) - \lambda_0(s)\right] p_0 \pi_{0,k}(s) +
      \left[\lambda_1(s) - \lambda_0(s)\right] p_1 \pi_{1,k}(s) \\
      & = &
      \left[\frac{\lambda_1(s)}{\lambda_0(s)} - 1\right]\lambda_0(s) p_0 \pi_{0,k}(s) 
      + 
      \left[1 - \frac{\lambda_0(s)}{\lambda_1(s)} \right]\lambda_1(s)
      p_1 \pi_{1,k}(s) 
  \end{eqnarray*}
  By plugging $x = \log(\lambda_1(s)/\lambda_0(s))$ and $x = \log(\lambda_0(s)/\lambda_1(s))$ intp the Taylor expansion $e^x \approx 1 + x$, we get 
  $$
  \left\{
  \begin{array}{c}
      \frac{\lambda_1(s)}{\lambda_0(s)} 
      \approx
      1 + \log\left(\frac{\lambda_1(s)}{\lambda_0(s)} \right) \\
      \frac{\lambda_0(s)}{\lambda_1(s)} 
      \approx
      1 - \log\left(\frac{\lambda_1(s)}{\lambda_0(s)} \right)
  \end{array}
  \right..
  $$
  If one plugs in the above Taylor expansion into $\mathcal A$, one can approximate $\mathcal A$ by
  \begin{eqnarray*}
      \mathcal A
      & \approx &
      \log\left(\frac{\lambda_1(s)}{\lambda_0(s)} \right) 
      p_0 \pi_{0,k}(s) \lambda_0(s) 
      +
      \log\left(\frac{\lambda_1(s)}{\lambda_0(s)} \right)
      p_1 \pi_{1,k}(s) \lambda_1(s) 
      \\
      & = &
      \log\left(\frac{\lambda_1(s)}{\lambda_0(s)} \right) \left[p_0 \pi_{0,k}(s) \lambda_0(s) + p_1 \pi_{1,k}(s) \lambda_1(s)\right]\\
      & = &
      \log\left(\frac{\lambda_1(s)}{\lambda_0(s)} \right) v'(s).
  \end{eqnarray*}
  So, we can simplify \eqref{equ: delta k simple} into
  $$
    \Delta_k 
    \approx
    \int_{0}^{\tau_k} w(s)
    \frac{
      p_{0,k} \pi_{0,k}(s) \;
      p_{1,k} \pi_{1,k}(s)
    }{\pi_k(s)^2}
    \log\left(\frac{\lambda_1(s)}{\lambda_0(s)} \right) v'(s) ds.
  $$
  By normalizing the weights of $\log\left(\frac{\lambda_1(s)}{\lambda_0(s)} \right)$, we have the conclusion in Section \ref{sec: design char}.

\section{Bridge the average HR from the AHR method and the WLR test}
\label{appendix: 2 beta bridge}
 
  Upon comparing $\varphi^{(\text{lr})}$ with $\varphi^{(\text{wlr})}$, it becomes evident that the two equations differ from each other. This difference is mainly due to their underlying assumptions. The $\varphi^{(\text{lr})}$ is derived from the piecewise model, as specified by Assumption \ref{assump: ahr piecewise}, whereas this assumption is not used in the derivation of $\varphi^{(\text{wlr})}$. If we introduce Assumption \ref{assump: ahr piecewise} into $\varphi^{(\text{wlr})}$, a mapping can be established to relate these two. Specifically, if we set
    $$
    w(s) = 
    \frac{
      \left(
      \frac{1}{1/d_{0,m}+1/d_{1,m}}
      \right)^{-1}
    }{
      p_{0, m} p_{1, m} d_m
      \sum_{i=1}^M
      \left(
        \frac{1}{1/d_{0,i}+1/d_{1,i}}
      \right)^{-1}
    }
  $$
  for any $s$ in the $m$-th interval for a generally $m = 2, \ldots, M$ in $\varphi^{(\text{lr})}$, then $\varphi^{(\text{wlr})}$ shares the same formula as $\varphi^{(\text{lr})}$ under the piecewise model (see Assumption \ref{assump: ahr piecewise}). Details to obtain this statement provided below.

  Notice the above $\Delta_k$ in \eqref{eq:WlrDletaK} takes the integration from 0 to the $k$-th analysis at time $\tau_k$. If it is at the end of the study, we have decomposed $\Delta_K$ -- via the piecewise model -- as
  \begin{eqnarray*}
    \Delta_K 
    & \approx &
    \sum_{\ell = 1}^M
    \int_{\tau_{\ell -1}}^{\tau_{\ell}} w(s)
    \frac{
      p_{0,\ell} \pi_{0,\ell}(s) \;
      p_{1,\ell} \pi_{1,\ell}(s)
    }{\pi_\ell(s)^2}
    \log\left(\frac{\lambda_1(s)}{\lambda_0(s)} \right) v'(s) ds\\
    & = &
    \sum_{\ell = 1}^M
    \int_{\tau_{\ell -1}}^{\tau_{\ell}} w(s)
    \frac{
      p_{0,\ell} \pi_{0,\ell}(s) \;
      p_{1,\ell} \pi_{1,\ell}(s)
    }{\pi_\ell(s)^2}
    \varphi_\ell \; v'(s) ds \\
    & = &
    \sum_{\ell = 1}^M
    \varphi_\ell\;
    \int_{\tau_{\ell -1}}^{\tau_{\ell}} w(s)
    \frac{
      p_{0,\ell} \pi_{0,\ell}(s) \;
      p_{1,\ell} \pi_{1,\ell}(s)
    }{\pi_\ell(s)^2}
     v'(s) ds
  \end{eqnarray*}
  If we further assume the dropout rate in the two arms is the same, then we have  $\pi_0(s) = \pi_1(s) = \pi(s)$ under the local alternatives \citep[Section 2.3][]{yung2020sample}. In this way, $\Delta_k$ can be simplified into 
  \begin{eqnarray*}
      \Delta_K 
      & \approx &
      \sum_{\ell = 1}^M
      \varphi_\ell \; p_{0,\ell}  \; p_{1,\ell}
      \int_{\tau_{\ell -1}}^{\tau_{\ell}} 
      w(s) \; v'(s) ds. 
  \end{eqnarray*}

  When 
  $
    w(s) = 
    \frac{
      \left(
      \frac{1}{1/d_{0,\ell}+1/d_{1,\ell}}
      \right)^{-1}
    }{
      p_{0, \ell} p_{1, \ell} d_\ell 
      \sum_{i=1}^M
      \left(
        \frac{1}{1/d_{0,i}+1/d_{1,i}}
      \right)^{-1}
    }
  $
  when $s \in [\tau_{\ell - 1}, \tau_\ell)$, then we have
  \begin{eqnarray*}
      \Delta_K 
      & \approx &
      \sum_{\ell = 1}^M
      \varphi_\ell \; p_{0,\ell}  \; p_{1,\ell}
      \int_{\tau_{\ell -1}}^{\tau_{\ell}} 
      \frac{
      \left(
      \frac{1}{1/d_{0,\ell}+1/d_{1,\ell}}
      \right)^{-1}
      }{
      p_{0, \ell} p_{1, \ell} d_\ell 
      \sum_{i=1}^M
      \left(
        \frac{1}{1/d_{0,i}+1/d_{1,i}}
      \right)^{-1}
      } \;
      v'(s) ds  \\
      & = &
      \sum_{\ell = 1}^M
      \varphi_\ell \; p_{0,\ell}  \; p_{1,\ell}
      \frac{
      \left(
      \frac{1}{1/d_{0,\ell}+1/d_{1,\ell}}
      \right)^{-1}
      }{
      p_{0, \ell} p_{1, \ell} d_\ell 
      \sum_{i=1}^M
      \left(
        \frac{1}{1/d_{0,i}+1/d_{1,i}}
      \right)^{-1}
      }
      \int_{\tau_{\ell -1}}^{\tau_{\ell}} 
      v'(s) ds \\
      & = &
      \sum_{\ell = 1}^M
      \varphi_\ell \; p_{0,\ell}  \; p_{1,\ell}
      \frac{
      \left(
      \frac{1}{1/d_{0,\ell}+1/d_{1,\ell}}
      \right)^{-1}
      }{
      p_{0, \ell} p_{1, \ell} d_\ell 
      \sum_{i=1}^M
      \left(
        \frac{1}{1/d_{0,i}+1/d_{1,i}}
      \right)^{-1}
      }
      [v(\tau_\ell) - v(\tau_{\ell - 1})] \\
      & = &
      \sum_{\ell = 1}^M
      \varphi_\ell \; p_{0,\ell}  \; p_{1,\ell}
      \frac{
      \left(
      \frac{1}{1/d_{0,\ell}+1/d_{1,\ell}}
      \right)^{-1}
      }{
      p_{0, \ell} p_{1, \ell} d_\ell 
      \sum_{i=1}^M
      \left(
        \frac{1}{1/d_{0,i}+1/d_{1,i}}
      \right)^{-1}
      }
      d_\ell \\
      & = &
      \sum_{\ell = 1}^M
      \varphi_\ell 
      \frac{
      \left(
      \frac{1}{1/d_{0,\ell}+1/d_{1,\ell}}
      \right)^{-1}
      }{
      \sum_{i=1}^M
      \left(
        \frac{1}{1/d_{0,i}+1/d_{1,i}}
      \right)^{-1}
      }
  \end{eqnarray*}
  where $d_\ell$ is the expected number of events at the $m$-th interval.
  The logarithm of AHR can be calculated after normalizing the weights in $\Delta_K$, i.e.,
  \begin{eqnarray*}
    \varphi^{(wlr)} 
    & \approx &
    \frac{
      \sum_{\ell = 1}^M
      \varphi_\ell 
      \frac{
      \left(
      \frac{1}{1/d_{0,\ell}+1/d_{1,\ell}}
      \right)^{-1}
      }{
      \sum_{i=1}^M
      \left(
        \frac{1}{1/d_{0,i}+1/d_{1,i}}
      \right)^{-1}
      }
    }{
      \sum_{\ell = 1}^M 
      \frac{
      \left(
      \frac{1}{1/d_{0,\ell}+1/d_{1,\ell}}
      \right)^{-1}
      }{
      \sum_{i=1}^M
      \left(
        \frac{1}{1/d_{0,i}+1/d_{1,i}}
      \right)^{-1}
      }
    } \\
    & = &
    \sum_{\ell = 1}^M
      \varphi_\ell 
      \frac{
      \left(
      \frac{1}{1/d_{0,\ell}+1/d_{1,\ell}}
      \right)^{-1}
      }{
      \sum_{i=1}^M
      \left(
        \frac{1}{1/d_{0,i}+1/d_{1,i}}
      \right)^{-1}
      }.
  \end{eqnarray*}
  This is the formula to derive $\varphi^{\text{(lr)}}$.

\section{Deriving the expected number of events in the AHR method}
\label{proof: ahr -- sample size and events}

The key count we consider is the expected events in each time interval.
Specifically, it is the number of subjects with events in the interval $(\tau_{m-1}, \tau_m]$, which is denoted as
$\bar n(\tau_{m-1}, \tau_m)$ for any $m = 1, \ldots, M$.
We focus on the expected value of $\bar{n}(\tau_{m-1}, \tau_m)$ due to its usefulness in computing an average hazard ratio under the piecewise model, which is calculated as

\begin{equation}
  E\left( \bar{n}(\tau_{m-1}, \tau_m) \right)
  =
  \int_0^{\tau - {\tau_{m-1}}}
  g(u)
  P
  \left(
    \tau_{m-1} < T \leq \min(\tau_m, \tau - u),
    \;
    T \leq C
  \right) du
  \label{eq:EnDef}
\end{equation}
Here the random variable $T > 0$ denotes the subject time of an individual until an event.
And random variable $C > 0$ denotes the subject time of an individual until loss-to-follow-up.
Please note that $T,C$ are defined by $\{\widetilde T_m, \widetilde C_m \}_{m = 1, \ldots, M}$, i.e.,
$$
  T 
  = 
  \sum_{m=1}^M 
  \min\{T_m, \tau_m, \tau_{m-1}\}
  \prod_{j=1}^{m-1} \mathbbm 1\{T_j > \tau_j - \tau_{j-1}\}
$$
$$
  C 
  = 
  \sum_{m=1}^M 
  \min\{C_m, \tau_m, \tau_{m-1}\}
  \prod_{j=1}^{m-1} \mathbbm 1\{C_j > \tau_j - \tau_{j-1}\}.
$$

The integration in \eqref{eq:EnDef} sums subjects enrolled before time $\tau - \tau_{m-1}$.
This is because, for a subject to be in the count $\bar{n}(\tau_{m-1}, \tau_m)$, they must be enrolled prior to time $\tau - \tau_{m-1}$.
By dividing the integration interval $\int_0^{\tau-\tau_{m-1}}$ into two sub-intervals, i.e., $\int_0^{\tau - \tau_m}$ and $\int_{\tau - \tau_m}^{\tau - \tau_{m-1}}$, we can simplify equation \eqref{eq:EnDef} as

\begin{eqnarray} 
  &&
  \nonumber
  E \left( \bar{n}(\tau_{m-1}, \tau_m) \right)\\
  & = &
  \nonumber
  \int_0^{\tau - \tau_m}
  g(u)
  P\left( \tau_{m-1} < T \leq \tau_m, T \leq C \right) du
  + \\
  &&
  \nonumber
  \int_{\tau - \tau_m}^{\tau - \tau_{m-1}}
  g(u)
  P\left(\tau_{m-1} < T \leq \tau - u, T \leq C \right) du \\
  & = &
  \label{eq:En}
  \underbrace{
  G(\tau - \tau_m)
  }_{\mathcal A}
  \underbrace{
  P \left( \tau_{m-1} < T \leq \tau_m, T \leq C \right)
  }_{\mathcal B}
  + \\
  &&
  \nonumber
  \underbrace{
  \int_{\tau - \tau_m}^{\tau - \tau_{m-1}}
  g(u)
  P \left( \tau_{m-1} < T \leq \tau - u, T \leq C \right) du}_{\mathcal C} 
\end{eqnarray}

\begin{itemize}
    \item For $\mathcal A$ in \eqref{eq:En}, it can be simplified into
    $$
      \mathcal A = G_{M+1-m} \triangleq G(\tau_{M+1-m}).
    $$
    This is because $g(u) = \gamma_j$ when $u \in ( \tau_{j-1}, \tau_j]$, one has
    $$
      G_j \triangleq G(\tau_j) = G_{j-1} + \gamma_j (\tau_j - \tau_{j-1})
    $$
    with $G_0 = 0$.
    
    \item For $\mathcal B$ in \eqref{eq:En}, it can be simplified into
    \begin{eqnarray*}
      \mathcal B
      \triangleq
      d_m
      =
      & = &
      \underbrace{
        P(\min\{T, C\} > \tau_{m-1})
      }_{Q_{m-1}}
      P(0 < T_m \leq \tau_m - \tau_{m-1}, T_m \leq C_m) \\
      & = &
      Q_{m-1} 
      \left(
        1-e^{-(\lambda_m + \eta_m)( \tau_m -  \tau_{m-1})}
      \right)
      \frac{\lambda_m}{\lambda_m + \eta_m}
    \end{eqnarray*}
    For $Q_{m-1}$, one has
    $$
      Q_{m-1} 
      = 
      \prod_{j=1}^{m-1}
      \underbrace{
      P(\min\{T_m, Y_m\} > \tau_{m-1})
      }_{q_m}
      = 
      \prod_{j=1}^{m-1}
      e^{-(\lambda_{m-1} + \eta_{m-1}) (\tau_{m-1} - \tau_{m-2})}.
    $$

    \item For $\mathcal C$ in \eqref{eq:En}, by transferring $u$ into $v = u-\tau + \tau_m$, it can be simplified as
    \begin{eqnarray*}
      \mathcal C 
      & = &
      \int_0^{\tau_m - \tau_{m-1}}
      g(v + \tau - \tau_m)
      P \left( \tau_{m-1} < T \leq \tau_m - v, T \leq C \right) dv \\
      & = &
      \gamma_{M+1-m}
      \int_0^{\tau_m - \tau_{m-1}}
      P \left( \tau_{m-1} < T \leq \tau_m - v, T \leq C \right) dv \\
      & = &
      \gamma_{M+1-m}
      P(\min\{T, C\} > \tau_{m-1})
      \int_0^{\tau_m - \tau_{m-1}}
      P \left( T_m \leq v, T_m \leq C_m \right) dv \\
      & = &
      \gamma_{M+1-m}
      Q_{m-1}
      \frac{\lambda_m}{\lambda_m + \eta_m}
      \int_0^{\tau_m - \tau_{m-1}}
      \left(
        1 - e^{-(\lambda_m + \eta_m)v} 
      \right)
      dv \\
      & = &
      \gamma_{M+1-m}
      Q_{m-1}
      \frac{\lambda_m}{\lambda_m + \eta_m}
      \left(
        \tau_m -\tau_{m-1}
        -
        \frac{1 - e^{-(\lambda_m + \eta_m)(\tau_m - \tau_{m-1})}}{\lambda_m + \eta_m}
      \right) \\
      & = &
      \gamma_{M+1-m}
      Q_{m-1}
      \frac{\lambda_m}{\lambda_m + \eta_m}
      \left(
        \tau_m -\tau_{m-1}
        -
        \frac{1 - q_m}{\lambda_m + \eta_m}
      \right) 
    \end{eqnarray*}
\end{itemize}

By combining $\mathcal A, \mathcal B, \mathcal C$ together, we can simplify \eqref{eq:En} as 
\begin{eqnarray*}
  E \left( \bar{n}(\tau_{m-1}, \tau_m) \right) 
  = 
  G_{M+1-m}
  d_m
  +
  \frac{\lambda_m Q_{m-1} \gamma_{M+1-m} }{\lambda_m + \eta_m}
  \left(
    \tau_m -\tau_{m-1}
    -
    \frac{1 - q_m}{\lambda_m + \eta_m}
  \right) 
\end{eqnarray*}

\section{Local alternative assumption}
\label{appendix: local alternative}

Notice that the asymptotic variance of $Z_k^{\text{(wlr)}}$ in Section \ref{sec: test -- wlr} is $\widetilde\sigma^2_{b,k} / \sigma_k^2$. And in the existing literature, there are multiple proposals to simplify it. 

\begin{itemize}
    \item A common assumption is called \textit{local alternative} \citep{schoenfeld1981asymptotic}. It assumes 
$
  \sup_{t < \tau} 
  |
  \log\left[ \lambda_1(t) / \lambda_0(t) \right] 
  |
  = O(n^{-1/2}),
$
and this assumption makes the asymptotic variance $\widetilde\sigma_{b,k}^2 / \sigma_k^2$ in Section \ref{sec: test -- wlr} equal to $1 + o(n^{-1/2})$. Thus, an alternative approximation for the large-sample distribution of $Z_{k}^{(\text{wlr})}$ is 
$$
  Z_{k}^{(\text{wlr})}
  \overset{d}{\to}
  N\left( \sqrt{n_k} \theta_k, 1 \right).
$$

  \item Another assumption is called \textit{fixed alternative}. An example of a fixed alternative is the PH. Under the fixed alternative, $\widetilde\sigma^2_b / \sigma_k^2$ and 1 may both serve as approximations for the large-sample variance of $Z_k^{\text{(wlr)}}$, but none of them are the limiting variance of $Z_k^{\text{(wlr)}}$. Consequently, there is no guarantee that one is always more accurate than the others. Additionally, the convergence in distribution for $Z_k^{\text{(wlr)}}$ itself requires the assumption of local alternatives, so we do not recommend using the fixed alternative.

  \item In the literature, we also find the existence of \textit{distant alternative}. This assumption lies in the ART module in Stata \citep{wei2018u, gottlieb2025differences}. Basically, it approximates the asymptotic variance of $U_k$ by simulations. In this paper, we use local alternatives.
\end{itemize}

\bibliographystyle{chicago}
\bibliography{referenceNPH}

\end{document}